\documentclass[iop, revtex4]{emulateapj}

\usepackage{epsfig}
\usepackage{amsmath}
\usepackage{natbib}
\usepackage{booktabs}
\usepackage{tabularx}
\usepackage{hyperref}
\usepackage{xcolor}

\def \ergs {{\rm ~erg~s}$^{-1}$}

\newcommand{\NH}{N_{\rm H}}

\DeclareRobustCommand{\ion}[2]{%
\relax\ifmmode
\ifx\testbx\f@series
{\mathbf{#1\,\mathsc{#2}}}\else
{\mathrm{#1\,\mathsc{#2}}}\fi
\else\textup{#1\,{\mdseries\textsc{#2}}}%
\fi}

\shorttitle{AGNfitter}
\shortauthors{Calistro Rivera et al.}

\begin{document}
\title{AGN\textit{\lowercase{fitter}}: A Bayesian MCMC Approach to \\Fitting Spectral Energy Distributions of AGN}

\author{Gabriela Calistro Rivera \altaffilmark{2,1}}
\author{Elisabeta Lusso \altaffilmark{3}}
\author{Joseph F. Hennawi \altaffilmark{1}}
\author{David W. Hogg \altaffilmark{4,5,6,1}}

\altaffiltext{1}{Max Planck Institute for Astronomy, K\"onigstuhl 17, D-69117 Heidelberg, Germany}
\altaffiltext{2}{Leiden Observatory, Leiden University, P. O. Box 9513, 2300
RA Leiden, The Netherlands}
\altaffiltext{3}{INAF Osservatorio Astrofisico di Arcetri, I-50125 Florence, Italy}
\altaffiltext{4}{Simons Center for Data Analysis, 160 Fifth Avenue, 7th 
oor, New York, NY 10010, USA}
\altaffiltext{5}{Center for Cosmology and Particle Physics, Department of Physics, New York University, 4 Washington Pl., room 424, New York, NY 10003, USA}
\altaffiltext{6}{Center for Data Science, New York University, 726 Broadway, 7th 
oor, New York, NY 10003, USA}

\email{calistro@strw.leidenuniv.nl}

\date{\today}

\begin{abstract}
\small
\noindent We present AGNfitter, a publicly available open-source algorithm implementing a
fully Bayesian Markov Chain Monte Carlo method to fit the spectral
energy distributions (SEDs) of active galactic nuclei (AGN) from the sub-mm to the UV, allowing
one to robustly disentangle the physical processes responsible for their
emission. AGNfitter makes use of a
large library of theoretical, empirical, and semi-empirical models to
characterize both the nuclear and host galaxy emission
simultaneously. The model consists of four physical emission
components: an accretion disk, a torus of AGN heated dust, stellar
populations, and cold dust in star forming regions. AGNfitter
determines the posterior distributions of numerous parameters that
govern the physics of AGN with a fully Bayesian treatment of errors
and parameter degeneracies,  allowing one to
infer integrated luminosities, dust attenuation parameters, stellar
masses, and star formation rates.
We tested AGNfitter's performace on real data by fitting the SEDs
of a sample of 714 X--ray selected
AGN from the XMM--COSMOS survey, spectroscopically classified
as Type1 (unobscured) and Type2 (obscured) AGN by their optical-UV emission lines.
We find that two independent model parameters, namely the reddening of the accretion
disk and the column density of the dusty torus, are good proxies for
AGN obscuration, allowing us to develop a strategy for classifying AGN
as Type1 or Type2, based solely on an SED-fitting analysis.
Our classification scheme is in excellent agreement with the spectroscopic classification, giving
a completeness fraction of $\sim 86\%$ and $\sim 70\%$, and an efficiency of $\sim
80\%$ and $\sim 77\%$, for Type1 and Type2 AGNs, respectively .
\end{abstract}

\section{Introduction}

Active galaxies host in their nuclei (AGN) the most efficient energy sources in the universe: accreting supermassive black holes (SMBHs) that convert significant fractions of the accreted material rest-mass energies
into powerful electromagnetic radiation. The evolution and properties of the host-galaxies of AGN are closely connected to the formation and growth of the SMBHs, as supported by both observational evidence \citep{1998AJ....115.2285M,2003ApJ...589L..21M} and cosmological simulations \citep{2005MNRAS.361..776S,2006ApJS..163....1H}.
The study of the SMBH/host-galaxy co-evolution demands a proper characterisation of their properties; for instance the intrinsic AGN
luminosity, the covering factor of nuclear obscuring medium, the stellar mass and star formation rate (SFR) of the host galaxies and the amount of this emission that is reprocessed by dust. These physical parameters are all encoded in the observed spectral energy distributions (SED) of the sources. In order to constrain these physical parameters and to place AGN into the context of galaxy evolution, it is fundamental to disentangle the contribution of the AGN from the host-galaxy in the observed SED.

The AGN SED covers the full electromagnetic spectrum from radio to X-rays. 
The most prominent features are the "infrared bump" at $\sim10-20$ $\mu$m,
and an upturn in the optical-UV, the so-called "big-blue bump"
(BBB; \citealt{1989ApJ...347...29S,1994ApJS...95....1E,2006ApJS..166..470R,2011ApJS..196....2S,2012ApJ...759....6E,2013ApJS..206....4K}).
The BBB is thought to represent the emission from the
accretion disk surrounding the SMBH, while the
mid-infrared
bump is due to the presence of dust that re-radiates a
fraction of the optical-UV disc photons at infrared wavelengths.
The presence of this screen of gas and dust surrounding the accretion
disc (dusty torus) is the foundation of the unified model for AGN
\citep{1993ARA&A..31..473A,1995PASP..107..803U}, which explains
the differences in their spectral characteristics as an effect of viewing angle with respect to a dusty obscuring torus. 
These spectral differences classify AGN into unobscured (Type1) and 
obscured (Type2) due to the presence or absence respectively 
of broad emission lines in their optical spectra. 
While in Type1 AGN the observer has a direct view into
the ionized gas clouds Doppler-broadened by the SMBH potential (the broad line region),
in Type2 AGN the line emission of these clouds is completely or partially extincted depending on the inclination angle respective to the torus, leaving only the narrow line emission to be observed.

An obvious complication in the study of the host galaxy properties in AGN is that the emission of the central nuclei outshines the galaxy light, therefore it becomes extremely difficult to derive constraints on the stellar populations. 
On the other hand, for obscured AGN the host-galaxy light may be the dominant component in the optical/near-infrared SED, making it challenging to estimate
the intrinsic nuclear power.

A common approach to tackle this problem is to fit the SED with different combinations of theoretical models and/or empirical templates for each individual emission component (eg. \cite{2008MNRAS.388.1595D}). 
The complexity of the models increases as a more accurate description of the underlying physics is needed. In this way, the number of unknown parameters increases as well, including the unavoidable existence of degeneracies and correlations among them. The best statistical approach for dealing with parameter degeneracies are Bayesian methods, which also allows the user to obtain reliable confidence ranges for parameter estimates.
The importance of a Bayesian study for general SED fitting has provoked the development of several Markov Chain Monte Carlo (MCMC)-based algorithms in the last years, principally in SED studies of quiescent galaxies (e.g. \cite{acquaviva11},  \cite{serra11}, \cite{pirzkal12}, \cite{johnson13}). 
These algorithms sample the space to infer galaxy parameters, which have been derived from stellar population synthesis models. 

Parallel AGN studies at single wavelength regimes have demonstrated that no monochromatic diagnostic can achieve a complete characterisation of AGNs \citep{juneau13} and combined multi-wavelength approaches are necessary for a comprehensive exploration of the physics of active galaxies \citep{lusso12}. To date, the publicly available SED fitting codes in the literature do not include any modelling of the broad-band AGN emission in a panchromatic approach, but have focused their analysis exclusively at infrared wavelengths (\citet{sajina06}, \cite{hanandhan12}, \citet{berta13} and \citet{hernan-caballero15}).

We have thus developed AGNfitter: a probabilistic SED-fitting tool based on Markov Chain Monte Carlo sampling. This code is designed to simultaneously disentangle the physical components of both AGN and host galaxy from optical-UV to sub-mm wavelengths and to infer the posterior
distribution of the parameters that govern them.
Our Bayesian method allows us to robustly perform this inference in order to recover the parameters with a complete description of their uncertainties and degeneracies through the calculation of their probability density functions (PDFs). The MCMC technique allows us to probe the shape of these PDFs, and the correlations among model parameters, giving far more information than just the best fit and the marginalized values for the parameters.
The efficiency and speed of the algorithm makes it capable of treating large samples of AGN and galaxies, enabling statistical studies of nuclear obscuration in the context of AGN classification. The current version of AGNfitter is publicly available on /github-link.

This paper is structured as follows. In \S \ref{sec:models} we describe the construction of the total AGN and host galaxy models used in the code. In \S \ref{sec:MCMC} we will explain the technical details of the code and the MCMC implementation used. For an illustration of the astrophysical capabilities of AGNfitter, we show in Section \ref{sec:mockdata} the results of the application of the code on synthetic data. In Sections \ref{sec:dataset} and \ref{sec:realdata} we describe the sample selection of Type1 and Type2 AGN of the XMM\textendash COSMOS survey and show the results of AGNfitter on this sample.

We adopt a concordance flat $\Lambda$-cosmology with $H_0=70\, \rm{km \,s^{-1}\, Mpc^{-1}}$, $\Omega_\mathrm{m}=0.3$, and $\Omega_\Lambda=0.7$ (\citet{komatsu09}, \citet{planck14}).

\section{The code AGNfitter}\label{sec:models}

AGNfitter produces independent samples of the parameter space to fit the observational data, which is constituted by photometric fluxes ranging from the optical-UV to the sub-mm.
Following \citet{lusso13} we consider four independently modelled components, which cover in total a rest-frame frequency range of log$\nu = 11-16$ ($\lambda = 0.01- 100 \mu$m). These models are specifically the accretion disk emission (BBB), the hot dust surrounding the accretion disk (the so-called torus), the stellar population of the host galaxy, and emission from cold dust in galactic star-forming regions. 
We make the assumption that the broad band SED can be constructed as a linear combination of these components with their relative contribution depending on the nature of the source. In order to properly mimic the observed fluxes, the emission templates F$_{\lambda}$ need to be integrated against the telescope response curves S($\lambda$) corresponding to the spectral bands used, following
\begin{equation}
F_{\rm S} = \dfrac{\int \lambda F_{\lambda}S(\lambda)\hspace{0.1cm}d\lambda}{\int \lambda S(\lambda)\hspace{0.1cm}d\lambda},
\end{equation}
where F$_{\rm S}$ is the resulting filtered flux.
Depending on the photometric data available, the corresponding filter curves can be chosen from the filter library included in AGNfitter, which is a compilation of COSMOS filters published in \url{http://www.astro.caltech.edu/~capak/filters/index.html}. New filters that are not available in this list can be easily added by the user.
\begin{figure*}[t]
\centering
\includegraphics[width=0.6\linewidth]{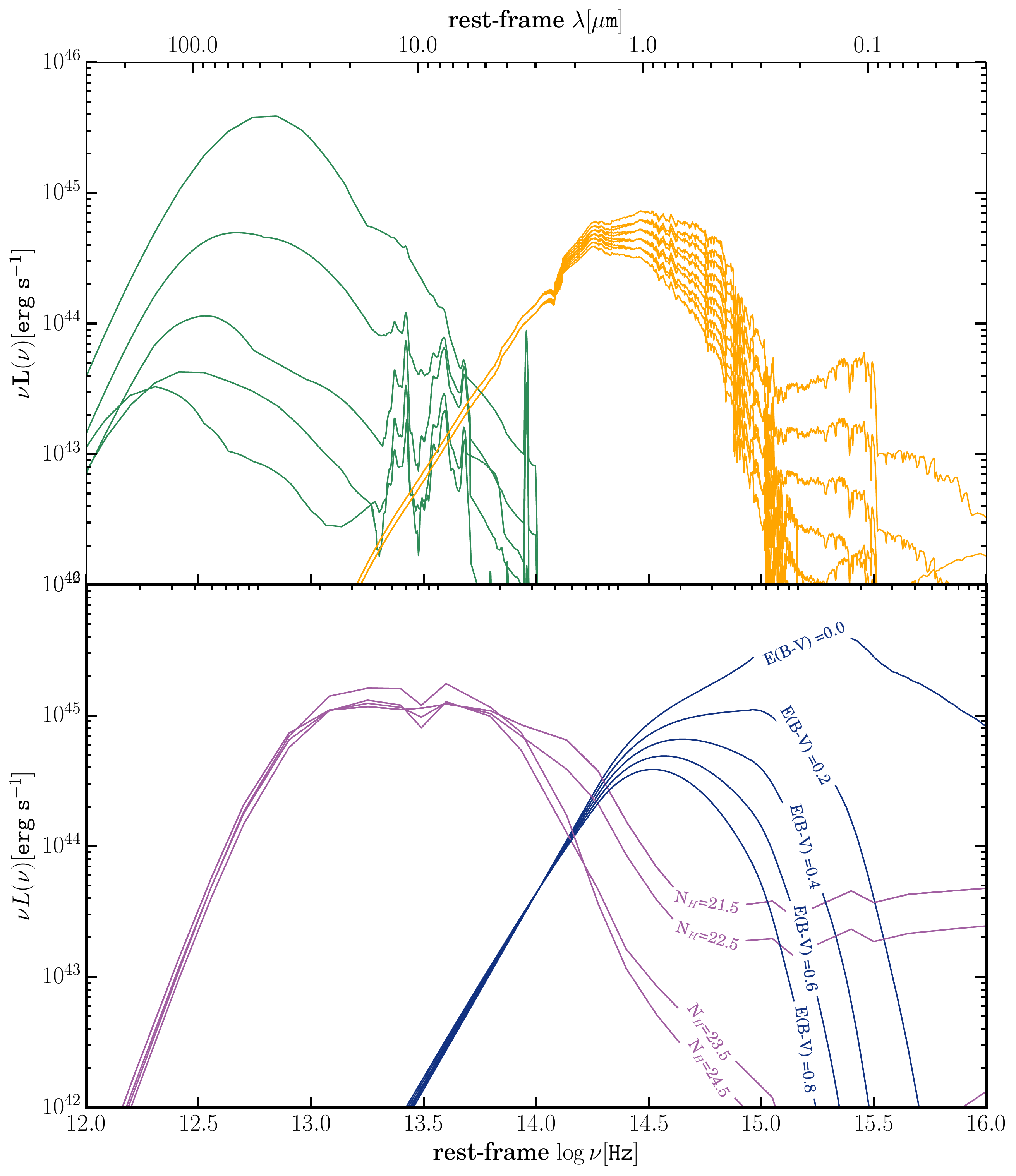}
\caption{Examples of templates employed in AGNfitter, which models AGN SEDs considering four different components. \textit{Upper panel}: The green and the orange solid lines correspond to different templates of the starburst and host-galaxy components, respectively.
\textit{Lower panel}: The purple and blue solid lines correspond to the hot-dust emission at different column density values log N$_{\rm H}$ and the BBB templates with increasing reddening $\rm E(B-V)_{bbb}$, respectively.}
\label{fig:templates}
\end{figure*}

\subsection{Accretion disk emission (Big Blue Bump)}

The most prominent feature in the AGN SED is the BBB, which is produced by the thermal radiation emitted by 
matter accreating into the central SMBH \citep{richstone80}. AGNfitter models this emission using a modified version of the empirical template constructed by \cite{richards06}.

This template is a composite spectrum of 259 Type1 QSO SEDs selected from the Sloan Digital Sky Survey (SDSS, York et al. 2000).
Given that the mid-infrared region of the SED will be modelled by independent warm dust templates in AGNfitter, we have neglected the near-infrared regime originally included in the Richards et al. SEDs. We extrapolated the BBB template from 1$\mu$m to longer wavelengths assuming $F_\nu\propto\nu^{-2}$ (i.e. Rayleigh-Jeans tail of a black body).
Our BBB component is modelled by a single template and the only adjustable parameter is a normalization factor called 'BB' (see Table~\ref{table:par}) determining its amplitude.

The emitted BBB spectrum can be altered by extinction from dust along the line of sight.
We model this effect by applying the \citet{prevot84} reddening law for the Small Magellanic Clouds (SMC, which seems to be appropriate for Type-1 AGN, \citealt{2004AJ....128.1112H,salvato09}) to the template of the emitted BBB SED in our model. 
Corresponding to the dust screen model, the extinguished flux $\rm f_{red}$ is given by
\begin{equation}
\rm f_{red}(\lambda)= f_{em}(\lambda) \times 10^{\rm -0.4 A_{\lambda}},
\end{equation}
where $ \rm f_{em}$ is the emitted flux. The value $\rm A_{\lambda}$ depends directly on the Prevot reddening law $k(\lambda)$ and the reddening parameter $\rm E(B-V)_{bbb}$, which is an input parameter in AGNfitter, following 
\begin{equation}
\rm A_{\lambda} = k(\lambda)  \rm E(B-V)_{bbb}.
\end{equation}
The exact function we use for the Prevot's law is 
\begin{equation}
k(\lambda) = 1.39 (10^{-4}\lambda)^{-1.2} - 0.38.
\label{eq:prevot}
\end{equation}
where $\lambda$ are in $\mu$m.

AGNfitter explores a discrete grid of reddening values in the range of $\rm 0\leq E(B-V)_{bbb}\leq 1$ in steps of 0.05 and uses equation~(\ref{eq:prevot})
to construct the reddening function. Since the MCMC requires a continuous parameter space, we overcome the discrete distribution of the reddening values through a k-Nearest Neighbour Interpolation (kNN), associating any $\rm E(B-V)_{\rm bbb}$ in the sampling to the nearest $\rm E(B-V)_{\rm bbb-grid}$ of our templates. 
A sub-sample of our BBB templates with different reddening levels is presented in Figure \ref{fig:templates} (blue solid lines).

\subsection{Hot dust emission (torus)}

The nuclear hot-dust SED models are taken from \citet{silva04}. 
These empirical templates were constructed from a large sample of Seyfert galaxies having reliable observations of the nuclear intrinsic NIR and MIR emission and are corrected by carefully removing any galaxy contribution.
The photometric data obtained from the Seyfert infrared observations were then interpolated based on a radiative transfer code GRASIL \citep{1998ApJ...509..103S} to obtain full AGN infrared SEDs.  This code simulates the emission of dust contents heated by a central source, which has a typical AGN intrinsic spectrum. 
The infrared SEDs are divided into 4 intervals of absorption: $\NH<10^{22}$ cm$^{-2}$ for Seyfert 1s (Sy1), $10^{22}<\NH<10^{23}$ cm$^{-2}$, $10^{23}<\NH<10^{24}$ cm$^{-2}$, and $\NH>10^{24}$ cm$^{-2}$ for Seyfert 2s (Sy2) . 
As explained in \citet{silva04}, the main difference between the SEDs as a function of $\NH$ is the increment of absorption in the near-IR at $\lambda < 2 \mu$m and some mild silicate absorption at 9.7 $\mu$m for larger $\NH$ values. A slight increase of the overall IR emission can be also observed at $\NH>10^{24}$ cm$^{-2}$, as a product of a higher covering factor of the circumnuclear dust at these $\NH$ values. In the context of the unified AGN model, the increasing $\NH$ values arise from different viewing angles with respect to the torus. We expect thus to observe Type1 AGNs of low $\NH$ values with hot dust emission at near-infrared wavelengths, while in the case of Type2 AGN this near-IR emission may be extincted by the dust distribution of high $\NH$ values.  The four templates employed in the code are plotted in Fig.~\ref{fig:templates} with the purple solid line.
In order to reduce the degree of discreteness in our parameter space due to the small number of torus templates, we performed a further interpolation in between these four templates. We then constructed an empirical function f($\NH$) which produces a finer parameter grid for the torus models. Our final SED grid consist on 80 templates corresponding to a range of log $\NH$ = [21,25] in intervals of $\Delta \NH = 0.05$. Also here, we overcome the discreteness of the $\NH$ parameter through a k-Nearest Neighbour Interpolation (kNN), associating any $\NH$ in the sampling to the nearest $\rm N_{\rm H-grid}$ of our templates. A further parameter adjusting this contribution is the corresponding amplitude parameter 'TO' as listed in Table \ref{table:par}.

\subsection{Stellar emission}\label{subsec:galaxymodel}

To construct the library of templates for the stellar emission of the host
galaxy we follow the standard approach and use the stellar population
synthesis models by \citet{bruzual03}. These models predict the
rest-frame flux of single stellar populations (SSPs) at different ages
and for different star formation rates. The evolution of the stellar
population is computed by assuming a constant metallicity, leaving
as free parameters both the age of the
galaxy ($age$) and the star-formation history time-scale ($\tau$).  

We have assumed a Chabrier (2003) initial mass function
(IMF) and an exponentially declining star formation history modulated
by the time scale $\tau$ as $\psi(age) \propto e^{-age/\tau}$. For the
purposes of this analysis a set of galaxy templates representative of
the entire galaxy population from passive to star forming is
selected. To this end, 10 exponentially decaying star formation
histories (SFHs) with characteristic times ranging from $\tau = 0.1$
to $10$\,Gyr and a model with constant star-formation rate are included.

For each SFH, a subsample of ages is selected, to avoid both
degeneracy among parameters and to speed up the computation. The grid of ages used to calculate the galaxy templates
covers ages from 0.2 to 11 Gyr in steps of $\log\,age \sim 0.1$. In
particular, early-type galaxies, characterized by a small amount of
ongoing star formation, are represented by models with values of
$\tau$ smaller than $1$ Gyr and ages larger than $2$\,Gyr,
whereas
more actively star forming galaxies are represented by models with
longer values of $\tau$ and a wider range of ages from $0.1$ to
$10$ Gyr.  An additional constraint on the age
implies that for each source, the age has to be smaller than the age
of the universe at the redshift of the source.

Altogether, 90 different stellar emission templates from the
\citet{bruzual03} models are included in AGNfitter. A small subsample
of these templates is shown in the upper panel of
Fig.~\ref{fig:templates} with the orange solid line.  Since the MCMC
requires a continuous parameter space, we overcome the discrete
distribution of the galaxy templates parameters through a k-Nearest
Neighbour Interpolation (kNN). In this way we choose from the grid the
nearest combination ($\tau_{grid}$, $age_{grid}$) corresponding to any
pair of parameter values ($\tau$, $age$).
Additionally, since stellar emission may be altered by extinction along the line of sight, the templates are reddened according to the  Milky Way-like reddening law derived in \citet{2000ApJ...533..682C}. 
The reddening values $\rm E(B-V)_{\rm gal}$ range between 0 and 0.5 with a step of 0.05. Including the extinction effect on the \citet{bruzual03} SED templates, the total AGNfitter stellar library is composed by 900 stellar templates. Finally, the amplitude of the stellar contribution on the total SED is adjusted by the normalizing parameter 'GA'. 

\subsection{Cold Dust Emission from Star-Forming Regions}\label{subsec:SBtemplates}

AGNfitter models the emission of cold dust in star forming regions using the semi-empirical starburst template libraries by \citet{chary01} and \citet{dale02}. These template libraries represent a wide range of SED shapes and luminosities and are widely used in the literature. 

The \citet{chary01} template library consists of 105 templates based on the SEDs of four prototypical starburst galaxies (Arp220 (ULIRG); NGC 6090 (LIRG); M82 (starburst); and M51 (normal star-forming galaxy)). They were derived using the \citet{1998ApJ...509..103S} models
with the mid-infrared region replaced with ISOCAM observations between 3 and 18$\mu$m (verifying that the observed values of these four galaxies were reproduced by the templates). These templates were then interpolated between the four prototypical starburst galaxies to generate a more diverse set of templates. 

The \citet{dale02} templates are updated versions of the widely used \citet{2001ApJ...549..215D} templates. These models involve three components, large dust grains in thermal equilibrium, small grains semistochastically heated, and stochastically heated PAHs. They are based on IRAS/ISO observations of 69 normal star-forming galaxies in the wavelength range 3\textendash100$\mu$m. \citet{dale02} improved upon these models at longer wavelengths using SCUBA observations of 114 galaxies from the Bright Galaxy Sample (BGS, see \citealt{1989AJ.....98..766S}), 228 galaxies observed with ISOLWS (52\textendash170$\mu$m; Brauher 2002), and 170$\mu$m observations for 115 galaxies from the ISOPHOT Serendipity Survey (\citealt{2000A&A...359..865S}). 

Altogether, the total far-infrared library used in AGNfitter is composed by 169 templates, which are parametrized by their different IR (8--1000$\mu$m) luminosities in the range IRlum$=10^8-10^{12}L_\odot$. The IRlum values were already pre-computed for each template in the original library provided by \citet{chary01} and \citet{dale02}. It is important to note, that the IRlum parameter is used in AGNfitter solely with the aim of indexing the templates in the library, hence it does not have any physical relevance, and it should not be confused with the computed total infrared luminosity of the sources (denoted as $\rm L_{\rm IR, 8-1000~\mu m}$, see \S~\ref{subsec:integratedlum}). Also here, we overcome the discrete parametrization of this component through a k-Nearest Neighbour Interpolation, associating any IRlum value required by the sampling to the nearest IRlum$_{\rm grid}$ of our templates. The amplitudes of this component's contributions are further adjusted by the normalization parameter 'SB'. A small subsample of starburst templates are plotted in Figure~\ref{fig:templates} with green solid lines.

\subsection{Parameter Space}

\begin{table*}
\caption{Parameter space in AGNfitter: Description of the parameters and their value ranges. The value of age$(z)$ is the limiting maximal age for a galaxy at redshift z, given by the age of the universe at that redshift according to the chosen cosmology.}

\hspace{-2cm}
\begin{center}
\begin{tabular}{ c| c | c | c }
\toprule
Component & Parameter $\rho_i$ & Description & Range  \\
\midrule \midrule
BBB emission &BB &BBB normalization&[0,20]\\
\midrule
BBB emission & E(B-V$)_{\rm bbb}$&BBB reddening &[0,1.0] \\
\midrule
torus emission &log N$_{\rm H}$ [log cm$^{-2}$] &torus column density& [21,25] \\
\midrule
torus emission &TO & torus normalization&[0,20]\\
\midrule
stellar emission &$\tau$[Gyr] & exponential SFH time scale & [0.1,3.0] \\
\midrule
stellar emission &log $age$ [log yr]& age of the galaxy & [$10^5$,age$(z)$] \\
\midrule
stellar emission & GA & galaxy normalization&[0,20]\\
\midrule
stellar emission &E(B-V)$_{\rm gal}$&galaxy reddening& [0,0.5]\\
\midrule
cold dust emission &log IRlum $[\log\,\rm L_{\odot}]$ &parametrization of starburst templates &[7,15]\\
\midrule
cold dust emission &SB &starburst normalization& [0,20]\\
\bottomrule
\end{tabular}
\end{center}
\label{table:par}
\end{table*}

The parameter space describing the total active galaxy emission in AGNfitter is 10-dimensional, as given by the number of parameters listed in Table \ref{table:par}. It is composed by two different types of parameters: (1) physical parameters, which determine the shape of the templates and hence represent non-linear dependencies, and (2) amplitude parameters, on which the model depends linearly and re-scale the contribution strength of each component. The ranges that we assume each parameters to reside in is listed in Table \ref{table:par}.

\section{The MCMC Algorithm}\label{sec:MCMC}

The Bayesian approach in SED-fitting implies that given the observed photometric data, the posterior probability of the parameters that constitute the model $\vec{\rho}$, is given by

\begin{equation}
P(\vec{\rho}|data) = \dfrac{P(data |\vec{\rho})P(\vec{\rho})}{P(data) },
\label{eq:bayes}
\end{equation}
where $\vec{\rho}$ is the ten dimensional vector that includes the
parameters listed in Table \ref{table:par}.  The function
$P(\vec{\rho})$ represents our prior knowledge on the distribution of
parameters previous to the collection of the data, while the factor $P(data)$ can be regarded as a
constant of proportionality, since it is independent of the
parameters. The function $P(data |\vec{\rho})$ is the likelihood of
the data being observed given the model $\mathcal{L}(data ,\vec{\rho})
\equiv P(data |\vec{\rho})$.

Provided that the individual photometric measurements are independent and their uncertainties are Gaussian distributed, the likelihood function is given by
\begin{equation}
\label{eq:likelihood}
\mathcal{L}(data,\vec{\rho})\propto \prod^{n}_{i=0} \exp \left[-\dfrac{[data_i - f(\vec{\rho}|\nu)]^2}{2 \sigma_i^2}\right],
\end{equation}
where  $f(\vec{\rho})$ is defined to be the total active galaxy model (linear combination of the four model templates presented in \S \ref{sec:models}). 
One of the main advantages of the Bayesian approach is the possibility of calculating the posterior PDFs of a specific subset of  parameters of interest, ignoring other
nuisance parameters, in a process referred to as marginalization. 
The PDF of a single parameter $\rho_i$ is then calculated by integrating over all possible values of the nuisance parameters, propagating their uncertainties into the final result on $P(\rho_i)$,
\begin{equation}
P(\rho_i) = \int ^{k\neq i} d\rho_1... d\rho_k  \hspace{0.1cm}\hspace{0.1cm} P(\vec{\rho}|data).  
\label{eq:PDFintegral}
\end{equation}

Though the Bayesian approach presents many advantages in theory, AGNfitter requires integrations of a high-dimensional parameter space with parameters related in non-linear functions and it can be extremely computationally intensive in practice. This can be overcome using numerical power, through informed sampling of the parameter space using MCMC algorithms.

Following the Bayesian assumption that unknown parameters can be treated as random variables, the MCMC algorithm produces independent samples of the parameters' posterior PDFs in an efficient way through a random walk in the parameter space. The sequence of visited points in the walk is called a \textit{chain}, where each chain is a Markov process. The 'Markov' property ensures that every Monte Carlo (random) step is dependent only on the last visited point having no memory of previous steps.

The random walk is initialised at a position $\rho_0$, which we describe in detail in \S \ref{subsec:initialization} for the case of AGNfitter. The transition steps between visited points will be decided from a proposal distribution $P$ and the chosen step will be accepted or rejected based on the comparison of the probability of the previous and next trial points. This probability is computed depending on the priors and the likelihood of the model at the current visited point, which results from the calculation of Eq. \ref{eq:likelihood}. The chosen proposal distribution $P$ is an important factor in the efficiency of the sampling, since it can lead the chain to regions of convergence or to make them diverge towards regions of mainly rejected steps producing heavily correlated samples.

\subsection{Emcee} 

The MCMC core embedded in AGNfitter is the Python package Emcee \citep{fm13}, which is a pure-python implementation, slightly modified from the affine-invariant MCMC ensemble sampler developed by Goodman \& Weare et al. (2010). 
In comparison to codes based of the Metropolis-Hastings (MH) algorithm \citep{metropolis53, hastings70}, one advantage of Emcee is its affine-invariance property, which implies the clever choice of a proposal distribution which is invariant under any linear transformation of the posterior distribution to sample.
This makes the proposal distribution independent on any possible covariances among the parameters, requiring the hand-tuning of only 1 or 2 parameters rather than $\sim \rm N^2$ for a traditional (MH) algorithm in an N-dimensional parameter space. The fact that the efficiency of Emcee is independent of the dimensionality of the model used is a clear advantage for treating a complex problem as the one tackled by AGNfitter. 

A further property of the Emcee implementation is that the parameter space is explored through a set of chains (\textit{walkers}) that evolve in parallel as an ensemble, rather than through a single chain. This implies that at each step each walker is randomly assigned to a partner walker, moving along lines which connect the single chains to each other. This allows two great features of the code. On one side this makes the exploration of the parameter space extremely efficient, since the proposal distribution of each current walker is based on the information gained by the rest of the chains. On the other side, Emcee used the parallel evolution of the chain to allow the distribution of the processes into parallel computing taking advantage of multi-core processors.
Finally, the simple Python interface of Emcee allows more flexibility in its use. A brief discussion comparing several Python statistical packages which supports our choice of Emcee can be found in \cite{vanderplas14}.

\subsection{Initialization and burn in steps} \label{subsec:initialization}

In AGNfitter all walkers start in a random position in a region of very small radius around the central values of the parameter ranges (see Table \ref{table:par}). 
Before running the MCMC sampling, which should return the right PDFs, the walkers need to arrive to the region of interest, around which the maximum likelihood should be located. There is a certain number of steps used for this aim, and the period needed for this initial convergence is called {\it burn in}. The burn in steps will be neglected for the purposes of
sampling the posterior probability distribution of the parameters. 


Achieving the convergence of the chains is an important condition for the correct sampling of the target distribution and different methods on judging convergence are discussed in the literature \citep[e.g.][]{gelman92, cowles96, lewis02}.
As explained in detail in \cite{fm13},  a good convergence diagnostic consists in measuring the autocorrelation function and more specifically, the integrated autocorrelation time of the chains.
The inverse of the autocorrelation time estimates the fraction of the total sample which has converged into the relevant regions, i.e. the effective fraction, which samples of the target distribution. 
As a feature of Emcee, AGNfitter includes the option to calculate the autocorrelation time using the Python module acor2.
The longer the autocorrelation time, the more samples that are needed during the burn-in face to ensure independent samples of the target density.
A good convergence test recommended for AGNfitter chains consists in verifying than the autocorrelation time for each parameter is at least a factor of $\sim$ 32 shorter than the chain length.


One recurring issue in parallel evolving chain-ensembles, is that some
of the chains can get trapped in some uninteresting local but not
global maxima.  In order to deal with this problem, we have
introduced multiple burn-in processes. 
The basic idea is that, after each single burn-in sampling, all
walkers are relocated to the point of the highest likelihood (global maximum) visited
during the last set, restarting a new sampling set and avoiding that
some of the walkers remain in regions of local maxima. This process is
iterated ideally until the region of interest (i.e. around the
location of the maximum likelihood value) is achieved.  Unfortunately,
there is no established theory for how long the burn in process should
be, and this must instead be determined empirically for the problem at
hand.  The number of burn-in sets, as well as the number of steps in
each burn in, can be set by the user. For the size of the AGNfitter
parameter space, we recommend at least two burn-in sets of 10,000
steps each.


\subsection{MCMC steps}

After the burn-in period all further samples are included in the calculations of the posterior PDFs. The
length needed for the sampling depends on several factors such as the size and dimension of
the parameter space, the quality of the data (number of data points,
error sizes), whether the convergence of the chains has been achieved
previously in the burn-in process, and the density of the sampling
desired.  For the latter, the number of sampling points used for the
distribution is not only determined by the number of MCMC steps but
also by the number of walkers used for the parameter space
exploration.

\begin{figure*}[t]
\includegraphics[width=1\linewidth]{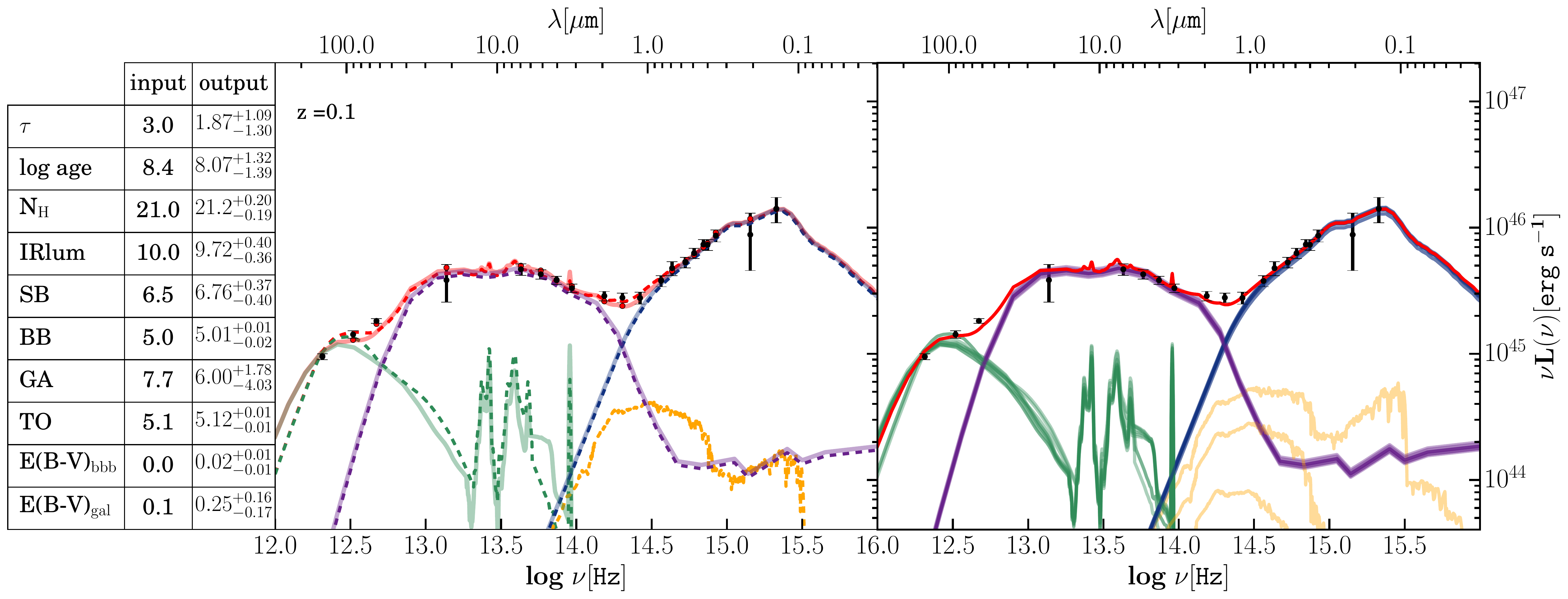}

\caption{SED output of a synthetic Type1 AGN: The colours of the component SED shapes are assigned similarly to Figure \ref{fig:templates}, while the linear combination of these, the 'total SED', is depicted as a red line. \textit{The table} provides the initial input values used for the construction of the mock AGN and the median values of the parameters' PDFs resulting from the fitting with the respective 16th and 84th percentiles. For the aim of comparison, we plot the SEDs as given by the input parameter values and the output best-fit values in the \textit{left panel}. The dashed lines are the SEDs corresponding to the input values and the solid transparent lines are the SED produced with the output best-fit parameters. In the \textit{right panel} we randomly pick eight different realizations from the posterior PDFs and over-plot the corresponding component SEDs in order to visualize the dynamic range of the parameters values included in the PDF. The total SEDs from the realizations are constructed and the mean is depicted as a red line.}

\label{fig:SED1}
\end{figure*}


\subsection{Running AGNfitter}

We set AGNfitter to use a number of 80 walkers and two burn-in sets each of 5,000 steps as default values. This can be freely changed based on the user's preferences. 
After the burn-in phase, the PDFs are sampled with a number of 10,000 steps during the MCMC procedure. For this default  configuration, the SED-fitting process takes in total about 5 minutes per source of computational time in a Macbook Pro with a 2.8 GHz Intel Core i5 processor, using only one core.
It should be noted that the time needed for sampling the PDFs is both source and data dependent, as the parameter space exploration time depends on how often the proposed steps are accepted or neglected. Since the consideration of each proposed step involves likelihood calculations, the time may increase for sources which are difficult to fit.

For the case of large catalogs, AGNfitter provides the possibility of fitting several sources simultaneously using the multiprocessing Python package, which distributes automatically several fitting processes into an specified number of computer cores. The availability of a multi-core computer would largely increase the efficiency of the code.

\subsection{Prior on the galaxy luminosity}
\label{Prior on the galaxy luminosity}

One of the problems we faced while fitting the optical-UV region of
the SED of Type 1 AGN was the degeneracy between the BBB and galaxies with very bright
and young stellar populations. A luminous Type 1 AGN ($L_{\rm bol}\sim10^{45-46}$ erg s$^{-1}$) can be either modelled with a BBB with zero reddening and negligible galaxy contribution, or with a highly reddened BBB (e.g., E(B--V)$\gtrsim0.5$) and a star-forming galaxy with significant contribution by the young stellar population. However, such galaxies will have exceptionally high
luminosities, much higher than typically observed.
Part of this degeneracy is due to lack of flexibility for the BBB template currently employed: the shape of the BBB (at zero reddening) is fixed, and it does not change with black hole mass\footnote{Our BBB template is representative of an optically selected AGN population with an average black hole mass of about $5\times10^8 ~M_\odot$ \citep{2006ApJS..166..470R}.}. On the other hand, stellar population synthesis models are much more flexible, since their shape depends on several parameters (e.g., $age$, $\tau$, metallicity, etc.).
To overcome this issue,  in AGNfitter we have implemented a simple prior on the galaxy luminosity based on the galaxy luminosity function.
Specifically, we considered the redshift evolution of galaxy Schechter luminosity functions. 
As estimated by \citet[see also \citealt{2010ApJ...708..505K}]{iovino10}, after fitting this redshift evolution, the characteristic absolute
magnitude in the B-band of the stellar population $M^\ast_B$ has the following functional form:
\begin{equation}
M^\ast_B = -20.3 -5 \log(h_{70}) - 1.1 \times z.
\label{eq:lumfct}
\end{equation}
$M^\ast_B$ includes a redshift evolution of roughly 1 magnitude between $z=0.1$ and $z = 1$ (see their \S~4 for further details), and it represents the characteristic absolute magnitude of a galaxy at a given redshift.
For each fit we computed the absolute magnitude from the galaxy template in
the B-band ($M_{\rm B,fit}$). This magnitude is then compared with the expected value from
equation (\ref{eq:lumfct}) at the source redshift. We set the likelihood to zero if $M_{\rm B,fit}$ is a factor of 10 times brighter than $M^\ast_{\rm B}$. 
 The choice of the threshold factor is rather arbitrary (and it can be changed by the user), but it provides a simple and effective way of preventing possible degeneracies between reddened BBBs and unphysically luminous stellar populations in Type 1 AGN.

\subsection{IGM absorption}

Significant amount of effort has been devoted in determining the AGN SED in the UV \citep{zheng97,2002ApJ...565..773T,2004ApJ...615..135S,2012ApJ...752..162S}. To estimate the {\it intrinsic} shape of the BBB one must take into account the intergalactic medium (IGM) absorption by neutral hydrogen along the line of sight \citep{2014ApJ...794...75S,2015MNRAS.449.4204L,2016ApJ...817...56T}.  Absorption from intergalactic \ion{H}{i}, blueward of \ion{Ly}{$\alpha$} emission in the AGN rest frame,  attenuates the source flux both in the Lyman series (creating the so-called \ion{H}{i} forest), and in the Lyman continuum at rest $\lambda_{\rm r}<912$\,\AA\ \citep[e.g.][]{1990A&A...228..299M}. Therefore, we should consider this effect while modelling AGN SEDs. 
To correct for the intervening \ion{Ly}{$\alpha$} forest and continuum absorption, one possibility is to employ a set of IGM transmission functions ($T_\lambda$) calibrated from multiple quasar absorption line observables (\citealt{2015MNRAS.449.4204L}, see also \citealt{2014ApJ...794...75S,2016ApJ...817...56T} for a different approach) over a wide range of redshifts. Such functions can be estimated statistically given the stochasticity of Lyman limit systems (e.g. \citealt{2011ApJ...728...23W,2014MNRAS.438..476P}) and critically depends on the parametrization of the \ion{H}{1} column density distribution \citep{1995ApJ...441...18M,2006MNRAS.365..807M,2014MNRAS.442.1805I,2014MNRAS.438..476P}.
Given that observations of the IGM are not present to constrain $T_\lambda$ at all redshifts, this implementation is problematic.
Although we recognise the importance of correcting broad-band photometry for IGM absorption, in order not to largely increase the model's complexity, we have been rather conservative and simply neglected all rest-frame data at wavelengths\footnote{For simplicity, we have considered the effective wavelength for each photometric band, neglecting their width.} shorter than 1250~\AA\ (i.e. $\log\nu/{\rm Hz}>15.38$). 

\section{Analysis of synthetic data}
\label{sec:mockdata}

\subsection{Synthetic SEDs}
\label{Synthetic SEDs}

In order to test the response of AGNfitter to AGN of different
obscuration properties, we construct two synthetic SEDs that mimic
typical Type1 and Type2 AGN.  These prototypes are built using four
model templates of known input parameters.  The parameter values are
chosen from representative values for the AGN populations and are
shown in tables in the left panel of Figure \ref{fig:SED1} for the
Type1 AGN and of Figure \ref{fig:SED2} for the Type2 AGN.  In the
middle panels of these figures, the four model templates that
correspond to those input parameter values are plotted as dashed lines
of different colors, where the colors correspond to different physical
components similar to Figure \ref{fig:templates}.  The red dashed line
is the total model calculated as a linear combination of these four
templates.  The total model SEDs are integrated against the broad band
filter curves that are included in the COSMOS photometry library to
produce mock photometric data points.  The uncertainties in the
photometry are mimicked through the addition of Gaussian random noise
of a standard deviation appropriate from typical uncertainties of each
band of the COSMOS photometry (see \S \ref{sec:dataset}). 
These mock photometric data are
plotted in the central and right panels as black dots with error bars.
\subsubsection*{MOCK Type1 AGN}

The mock Type1 AGN is constructed using the input parameter values specified in the Table of Figure \ref{fig:SED1} .
Since the optical-UV regime of the prototypical Type1 SED is dominated by the direct accretion disk emission, 
we choose a reddening value of $\rm E(B-V)_{\rm bbb}$= 0, following the observation that Type1 AGN  do not exhibit large amounts of extinction at these wavelengths (\cite{lusso12}, \cite{merloni14}). 
As obscuring media is not abundant in Type1s, the mid-IR/sub-mm regime is mimicked by a starburst (cold dust) and a torus template (warm dust) normalized to lower luminosities.  
We choose these luminosities accordingly to the obscuring fraction for Type1 AGN in XMM-COSMOS, which has been observed to approximately follow the relation  $\rm L_{\rm tor}$ = 0.5$\times$ L$_{\rm bbb}$ \citep{lusso13, merloni14}.
Considering the low optical depth medium observed for Type1 tori
 \citep{hatziminaoglou08}, the torus template is chosen to have the minimum column density of $\log \rm N_{\rm{H}} = 21.0$.
The galaxy template chosen is determined by parameter values of $\tau = 3$ and $\log age= 8.4$ ($age = 0.251~\rm{ Gyrs}$).
Because of the small stellar contribution to the total emission in Type1 AGNs, the values assumed for the $age$ and $\tau$ parameters do not play a significant role in modelling the shape of the total SED. 
Following these conditions we have modelled the MOCK Type 1 to be a redshift 0.1 bright, unobscured QSO, with direct
BBB luminosity $L_{\rm bbb, 1-0.1 \mu m} = 2.0 \times 10^{46}$ \ergs and torus luminosity $L_{\rm tor,1- 40\mu m } = 1.0 \times 10^{46}\rm $\ergs , where these luminosities are calculated by integrating over the wavelength ranges specified in the subscripts.

\subsubsection*{MOCK Type2 AGN}

\begin{figure*}[t]
\includegraphics[width=1\linewidth]{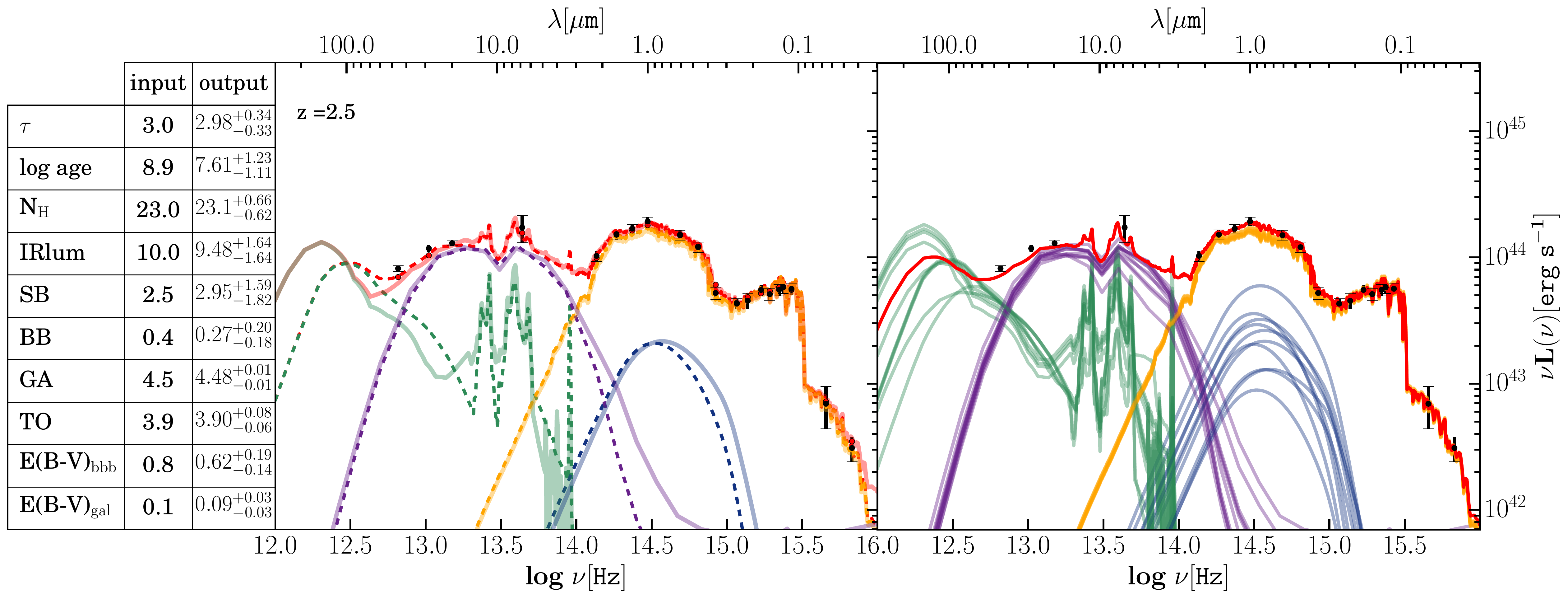}
\caption{SED output of a synthetic Type2 AGN: The colours of the component SED shapes are assigned similarly to Figure \ref{fig:templates}, while the linear combination of these, the 'total SED', is depicted as a red line. \textit{The table} provides the initial input values used for the construction of the mock AGN and the median values of the parameters' PDFs resulting from the fitting with the respective 16th and 84th percentiles. For the sake of comparison, we plot the SEDs as given by the input parameter values and the output best-fit values in the \textit{left panel}. The dashed lines are the SEDs corresponding to the input values and the solid transparent lines are the SED produced with the output best-fit parameters. In the \textit{right panel} we randomly pick eight different realizations from the posterior PDFs and over-plot the corresponding component SEDs in order to visualize the dynamic range of the parameters values included in the PDF. The total SEDs from the realizations are constructed and the mean is depicted as a red line.}
\label{fig:SED2}
\end{figure*}


The synthetic Type2 AGN is constructed following the definition that Type2s lack or have low accretion disk emission in the optical/UV, since this is obscured on nuclear scales. 
Therefore two properties have to be accounted for: (1) The dominant contribution to the total optical/UV luminosity is the stellar emission of the host galaxy, since the AGN contribution (BBB) is obscured. (2) The AGN emission obscured by the hot dust of the torus is reemitted in the IR, and thus the torus emission is expected to represent a large fraction of the accretion disk luminosity in a Type1 AGN (according to the AGN unified model\citep{1993ARA&A..31..473A,1995PASP..107..803U}). 

Following condition (1), the BBB template is normalized to a lower luminosity than the Type1 example, producing a total observed luminosity of $\rm L_{\rm bbb-red, 1-0.1 \mu m} =  4.4 \times 10^{43}$ \ergs after being extincted according to a reddening parameter of $\rm E(B-V)_{bbb}= 0.2$. If we correct for dust extinction, the mock Type 2 AGN present a dereddened BBB luminosity of $\rm L_{\rm bbb-dered, 1-0.1 \mu m} =  1.59\times 10^{44}$ \ergs.   
The dominant contribution of the stellar emission is chosen to be $\rm L_{\rm gal, 1-0.1 \mu m} = 1.9 \times 10^{44}$ \ergs corresponding to $\sim 81 \%$ of the total optical/UV contribution.
The host galaxy chosen for the mock Type 2 is a galaxy of regular stellar mass $ \rm M_{*} = 5.77 \times 10^{10} \rm M_{\odot}$, intermediate age ($\log~age = 8.9, age = 0.79~\rm{Gyrs}$) and following an exponential SFH with $\tau = 3.0$. 

Following condition (2), we model the mock Type 2 torus emission to have the luminosity $\rm L_{\rm tor, 1-40 \mu m} = 7.0 \times 10^{43}$ \ergs, which corresponds to about $52\%$ of the total MIR luminosity. 
The chosen template represents a torus of large column density ($\log \NH = 23$), which includes dust self-absorption, meaning that the emission from the inner, hotter part of the torus, emitting at shorter wavelength is highly absorbed (eg. \cite{1998ApJ...509..103S}, \cite{hatziminaoglou08}).

Following these conditions, we have modelled the mock Type2 AGN as a redshift $\rm z=2.5$ obscured active galaxy, with an intrinsic luminosity corresponding to $\rm L_{\rm bbb-dered, 1-0.1 \mu m} =  1.59\times 10^{44}$ \ergs.

\subsection{Recovery of parameters}
\label{subsec:recovery}

After running the code, the comparison between the columns in the tables of Figure \ref{fig:SED1} and Figure \ref{fig:SED2} shows a close agreement between the input and output values of the fitting parameters within the uncertainty ranges. This shows clearly the capability of AGNfitter in recovering accurate parameter values as well as 
error estimates which encompasses both the noise of the data and the degeneracies among the model parameters.

This can also be seen visually 
in the middle panels of Figures \ref{fig:SED1} and \ref{fig:SED2}
 where the dashed lines
represent the input SEDs used to create the synthetic data points and
the solid lines represent the output best-fit SEDs. As shown here,
both curves overlap almost completely, showing excellent agreement
between the input and output SEDs. Here, we have chosen to plot the
best-fit result as output SED, which is the SED given by the output
parameter values of highest likelihood. This result is comparable to
the output obtained through other customary SED-fitting tools that use
the $\chi^2$-minimization method. In both examples of Figures
\ref{fig:SED1} and \ref{fig:SED2} the best fit solution recovers
almost perfectly the input SED shape of each AGN model component.


\subsection{Posterior PDFs}

\begin{figure*}
  \centering
  \includegraphics[width=1\linewidth]{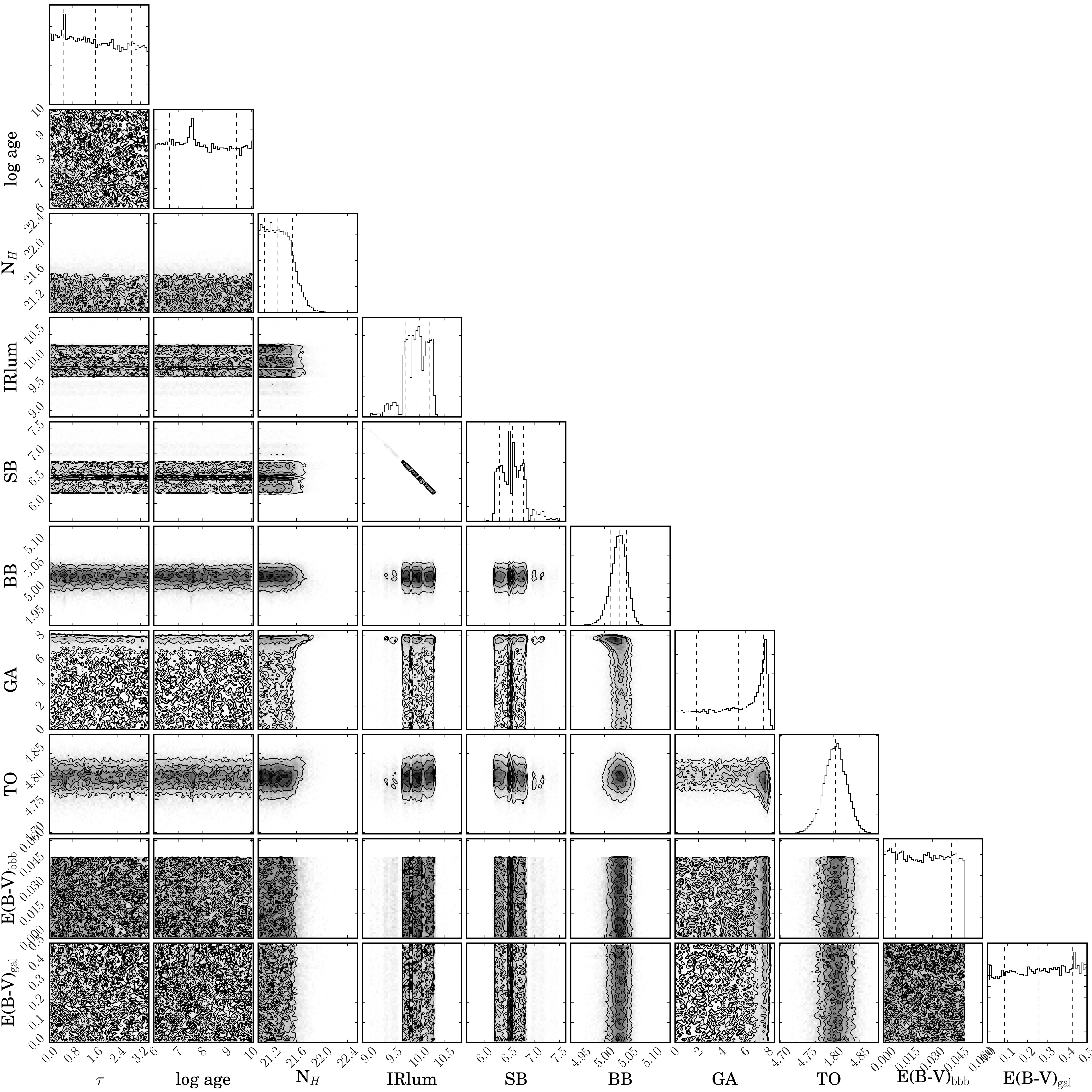}
  \caption{PDF triangle of the synthetic Type 1 AGN shown in Figure~\ref{fig:SED1}: 1- and 2 dimensional PDFs of the components of the parameter space explored by AGNfitter. The outer part of the triangle shows the 1-dimensional PDFs for each parameter as histograms. The lines indicate the medium and 1 $\sigma$ uncertainties. The inner part of the triangle shows the 2-dimensional PDFs for pairs of parameters as contours. The different contours shades represent the 0.5, 1, 1.5 and 2 $\sigma$.}
  \label{fig:pdftriangle}
\end{figure*}

One of the primary quality assessments products of the algorithm are the PDFs of the fitting parameters.
These are summarized in the PDF-triangle of each source and contain a wealth of information on the parameter dependencies.

In Figure \ref{fig:pdftriangle} the PDF triangle produced from the fitting of the synthetic Type1 AGN is presented as an example.
One-dimensional and two-dimensional PDFs of all parameters are depicted in the border and interior parts of the diagram, respectively.
The median value (50th percentile) and 16th and 84th percentiles of the PDF are reported by the dashed black lines. 
The two-dimensional PDFs are shown as probability surfaces, with contours corresponding to 38th, 68th, 88th and 95th percentiles (equivalent to 0.5, 1, 1.5 and 2 $\sigma$ for Gaussian distributions). 

The shapes of the 1-D and 2-D PDFs are a good indicator of the quality of the inference of a given parameter and hints to the existence of covariances among parameters. 
As can be noticed for the mock Type1 AGN in Figure \ref{fig:SED1}, parameters associated with the galaxy contribution ($\tau$, $age$, GA, E(B-V)$_{\rm gal}$) have very noisy and nearly flat PDFs. This is a consequence of the lack of recognisable galaxy features in the total SED of a Type1 AGN (total SED: red line in \ref{fig:SED1}) and thus, the stellar properties cannot be very accurately constrained. 
Nonetheless an upper limit for the normalisation parameter of this template (parameter GA) is well defined and a maximum contribution to the total luminosity can be infered.
Parameters determining the AGN contributions ($\rm \log N_{\rm H} ,BB, TO, \rm E(B-V)$) on the contrary show more defined posterior distributions and the resulting parameter values, as given by the percentiles, present smaller uncertainties.

As mentioned before, we interpolate between discrete model templates using Nearest-Neighbor interpolations in order to have a continuous parameter space. 
This has also an effect on the 1-dimensional PDFs of discrete parameters (in this example IRlum and E(B-V)$_{\rm bbb}$
in Figure \ref{fig:pdftriangle}), which sometimes show abrupt cut-offs on the regions outside the nearest neighborhoods of the discrete values chosen (in this example: $\rm IRlum=10$ and $\rm E(B-V)_{\rm bbb}=0$). 

The 2-dimensional PDF allows the recognition of covariances of the parameters. This can be clearly observed for instance in Figure \ref{fig:pdftriangle} for the parameters IRlum and SB. This strong covariance between these parameters occurs by construction, since as explained in \S~\ref{subsec:SBtemplates}
both the parameters SB and IRlum are normalization parameters of the same model. While SB is a normalization parameter introduced in AGNfitter, IRlum was predefined by the authors of the models to parametrize the SED shapes of the cold dust library based on their total luminosities.

\begin{figure}
  \centering
  \includegraphics[width=1\linewidth]{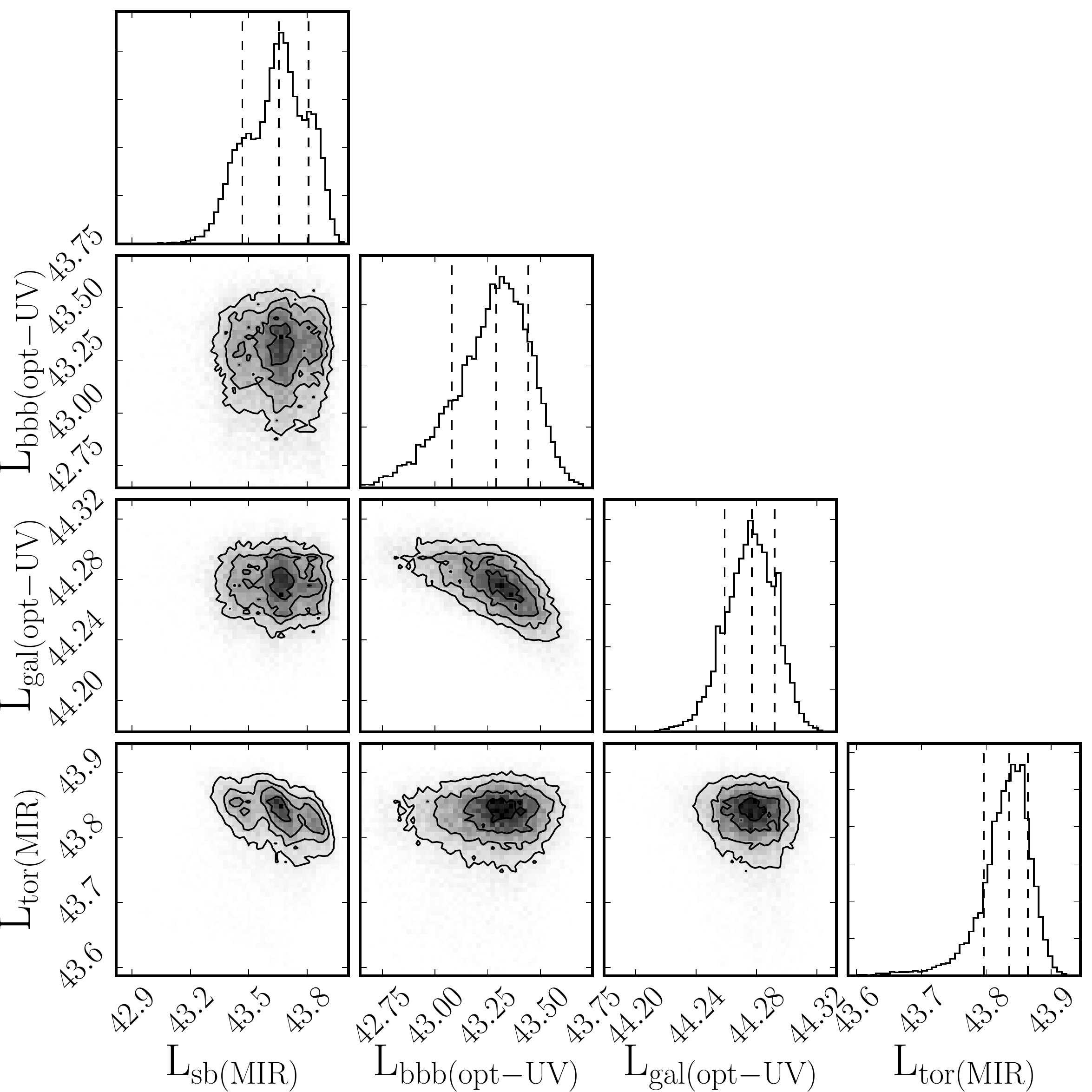}
  \caption{PDF triangle of the synthetic Type2 AGN shown in Figure~\ref{fig:SED2}: 1- and 2 dimensional PDFs of the default AGNfitter output on integrated luminosities is presented. The outer part of the triangle shows the 1-dimensional PDFs for each parameter as histograms. The lines indicate the medium and 1 $\sigma$ uncertainties. The inner part of the triangle shows the 2-dimensional PDFs for pairs of parameters as contours. The different contours shades represent the 0.5, 1, 1.5 and 2 $\sigma$. The 2D relative posterior of the integrated luminosities $L_{\rm bbb, ~0.1-1~ \mu m}$ and $L_{\rm gal,~0.1-1~ \mu m} $ show slightly twisted banana shaped contours, revealing that the estimation of the two parameters are degenerate. }
  \label{fig:lum_triangle_type2}
\end{figure}

Similarly, complex shaped 2D posterior distributions can result from
non-linear dependencies on model parameters.  As an example, Figure
\ref{fig:lum_triangle_type2} presents the PDF triangle of the
integrated luminosities of the synthetic Type2 AGN.
The 2D 
posterior 
of the integrated luminosities $L_{\rm bbb, ~0.1-1~ \mu m}$ and
$L_{\rm gal,~0.1-1~ \mu m} $ show slightly twisted banana shaped
contours, revealing that the estimation of the two parameters are
degenerate.  This can be easily understood observing the physical
components from which these luminosities are computed in the right
panel of Figure \ref{fig:SED2}.  Even though the galaxy templates
dominates the SED, both the reddened BBB and the galaxy template peak
around $1 \mu m$, such that both can contribute to the observed SED
around this wavelength. This naturally results in a slight degeneracy
in their respective integrated luminosities. See \S
\ref{subsec:SEDs} for a more detailed description of the SED plot.

One of the main advantages in estimating the parameters' PDFs is that robust uncertainties on the parameter inference can be now calculated. 
As describe above, AGNfitter estimates credible intervals from the PDFs (50th, 16th and 84th percentiles). 
Following our Bayesian approach we quote only these uncertainty values from now on in this paper. 
If requested by the user, AGNfitter returns as well the best-fit values for the parameters, which should be the one with the highest likelihood computed in the chain
in the case of convergence. Nonetheless, if the posterior distribution inferred by the sampling is not symmetric, the best fit value will differ from the median value.

All in all, the true values used for the construction of the mock Type-1 and Type-2 AGN  are robustly recovered by the median values and uncertainties of the parameters posteriors estimated by AGNfitter. 

\subsection{Integrated luminosities}
\label{subsec:integratedlum}

Integrated luminosities are estimated in order to probe the total power produced by a given radiative process. AGNfitter presents a set of functions, which can be easily adjusted by the user to obtain the preferred integration ranges. In the default version of AGNfitter we calculate four integrated luminosities over frequency ranges, which are key to the decomposition of host galaxy/AGN contributions ($L_{\rm bbb},L_{\rm gal},L_{\rm sb},L_{\rm tor}$), as well as two integrated luminosities which can be converted to key parameters for the physical description of the sources ($L_{\rm bol}$, $L_{\rm IR}$). Consistent with our Bayesian approach, AGNfitter computes the total posterior PDFs of the integrated luminosities and returns their median values and percentiles as representative values. Since integration algorithms are computationally expensive, the chains can be additionally 'thinned' (e.g. by a factor k), which means we discard all but every k-th observation with the goal of reducing the computational time.

$L_{\rm bbb, ~0.1-1~ \mu m}$, $L_{\rm gal,~0.1-1~ \mu m}$: 
We calculate the integrated accretion disk luminosity (BBB) and the integrated host galaxy emission in the frequency range from 0.1-1 $\mu$m, in the $\log \nu$ - $\log (\nu \rm L_{\nu})$ frame. This interval is chosen due to the simultaneous emission of AGN and host galaxy, which allows the study of their relative contribution.

$L_{\rm tor,~1-40 ~\mu m}$, $L_{\rm sb,~1-40 \mu m}$: 
Several studies have shown that the mid-infrared regime offers a good laboratory for the study of the ratio of AGN and galaxy contribution to total luminosity \citep[e.g.][]{lacy04, stern05,donley12,ciesla15}.
AGNfitter computes MIR-luminosities for both the torus and cold dust (starburst) emission components integrating within the relevant wavelength range ($1-40 \mu\rm m$).

$L_{\rm bol, ~0.1 \mu m - 1KeV}$: 
An important probe for the intrinsic power of an AGN is its bolometric
luminosity, which we calculate integrating the rest-frame AGN direct emission (BBB) within the frequency interval from 0.1 $\mu$m - 1 keV , in the $\log \nu$ - $\log (\nu \rm L_{\nu}$) frame.
We only consider the BBB
template and omit the torus template for the integration, since nearly
all photons emitted at infrared wavelengths are reprocessed
optical/UV/soft X-ray photons emitted into other directions and
re-radiated isotropically. Thus, adding the torus emission to the
L$_{\rm bbb}$ calculation would amount to double counting these
contributions to the luminosity (e.g. \citet{marconi04,lusso12}).
The X-ray emission would also contribute $\sim$20$\%$ to the total bolometric
budget but we omit this contribution since the fitting of X-ray data is not currently
implemented in AGNfitter.

$L_{\rm IR,~ 8-1000~ \mu m}$:
After correcting for any torus contamination, the 8-1000 $\mu$m luminosity is calculated integrating the cold dust emission (starburst component emission). This infrared luminosity is used to estimate the $\rm SFR_{FIR}$ as explained in detail in \S \ref{subsec:galaxyproperties}.

\subsection{Host galaxy properties} \label{subsec:galaxyproperties}

AGNfitter estimates galaxy properties such as stellar masses and star formation rates (SFR) and calculates the total posterior PDFs for these parameters. As before, the median values and uncertainties (16th and 84th percentiles) are given as representative values.
The total stellar masses are computed from the masses of the single stellar population (SSP) that determine each galaxy template. The normalization of the templates give us a constraint about the number of SSPs and in turn the total stellar mass producing the galactic optical emission.

AGNfitter makes two independent SFR diagnoses both in the optical/UV and the far-infrared (FIR).
In the optical regime, the normalization log parameter for the galaxy, $GA$, permits to scale the emission of the generic galaxy template to the total emission needed for the fit.
With the inferred total stellar mass and taking into account the age and the timescale of star formation history of the stellar populations ($\tau$), we can calculate the star formation rates according to:
\begin{equation}
\left(\dfrac{\rm SFR_{\rm opt/UV}}{M_{\odot} yr^{-1}} \right) = 10^{\rm{GA}}  \times \left(\dfrac{M_{\rm \ast sp}}{M_{\odot}} \right) \times \left( \dfrac{ \exp(-age/\tau)}{\tau~ yr^{-1}} \right) ,
\end{equation} 
where $M_{\rm \ast sp}$ is the mass of the stellar population.

The total FIR emission provides a different measurement of the SFR since it proves the emission of star forming regions reprocessed by dust grains in their surroundings. This estimation is of special relevance at redshifts 1 $\leq$ z $\leq$ 3, where the highest star-formation rate galaxies 
appear to be dominated by dusty starbursts with high infrared luminosities (\cite{chary01}, \cite{bouwens09}). In order to derive the SFR from the starburst model component that dominates the FIR, we use the calibration derived in \cite{murphy11} for the total infrared wavelength range (8-1000 $\mu$m),

\begin{equation}
\left(\dfrac{\rm SFR_{\rm IR}}{M_{\odot}yr^{-1}}\right) = 3.88 \times 10^{-44} \left(\dfrac{\rm L_{\rm IR}}{{\rm ~erg~s}^{-1}}\right).
\end{equation} 
Here, $L_{IR}$ is the luminosity integrated over the starburst template in the wavelength range noted above, after a proper substraction of the AGN torus contribution from the total IR luminosity.

AGNfitter does not include the assumption of energy balance between the direct stellar emission in the optical and the reprocessed emission by cold and warm dust in the IR. This is chosen consciously in order to prevent mixing-up possible data uncertainties of the optical and the FIR. 
Both components are thus computed independently and this allows to test the power of each wavelength regime in tracing the SFR. A good agreement between both diagnoses is expected, although the far-infrared SFR tracer is highly dependent on the availability of good FIR data.  In the cases where the data set does not contain enough FIR detections the MIR emission is associated entirely to the torus component. In this way, the starburst emission, and consequently the SFR$_{\rm FIR}$, are highly underestimated showing no agreement with the optical tracer.

\subsection{Spectral energy distributions}
\label{subsec:SEDs}

The probabilistic approach of AGNfitter can be better visualized through the plotting of several different SED realizations, constructed from values randomly picked from the full posterior PDFs of the parameters.
In the right panels of Figures \ref{fig:SED1} and \ref{fig:SED2}, a representative number of SED realizations are shown simultaneously to exemplify the dynamic range of individual template models and total SEDs contained in the probability distributions. 
The number of displayed realizations is 8 by default but can be freely chosen by the user. 
Each realization comprises a set of five lines of different colors; one red line that represents the total SED as the linear combination of the physical components, and four lines of different colors, which correspond to the different physical components, following the color coding of Figure \ref{fig:templates}.
Through the variety of SED shapes among single components it is easy to recognize which components are described by parameters with larger uncertainties.
This is clear from the output-SED fitted for the mock Type2 AGN in Figure \ref{fig:SED2}. 
There the BBB SED appears to be subject to large uncertainties due to a degeneracy among parameters determining the BBB amplitude (BB) and shape (altered by extinction parameter $\rm E(B-V)_{\rm bbb}$) and the galaxy normalization parameter.
Similarly, the parameters determining the starburst template and torus appear to be slightly degenerate producing a large dynamic range of SED shapes for these components.
The degeneracy among these parameters was discussed previously in the context of Figure \ref{fig:lum_triangle_type2}, where the 2D-PDFs of the parameters for the same source were introduced. 
The visualization of different SED shapes from PDF realizations stresses the relevance of a careful consideration of the parameter uncertainties with the aim of a proper physical interpretation of the SED decomposition.

\section{A test sample: XMM\textendash COSMOS}
\label{sec:dataset}

AGNfitter is a multi-purpose SED fitting tool. It can be applied to
different kinds of sources such as quiescent or starburst galaxies,
as well as AGN/quasars, given a source redshift and photometric data. In the rest of this
paper, we test the performances of AGNfitter by applying it to both 
obscured and unobscured X-ray selected AGN from the XMM\textendash COSMOS survey,
taking advantage of the unique multiwavelength coverage of the
COSMOS field \citep{scoville07}.

The XMM-COSMOS catalog comprises 1822 point-like X-ray sources detected by XMM-\textit{Newton} over an area of $\sim 2\,\rm deg^2$ for a total of $\sim 1.5$ Ms at a fairly homogeneous depth of $\sim 50$ ks (\citealt{hasinger07}, \citealt{cappelluti09}).
All the details about the catalog are reported in \cite{brusa10}.
We consider in this analysis 1577 X\textendash ray selected sources for which a reliable optical counterpart can be associated (see discussion in \citealt{brusa10}, Table 1) and we restrict our sample to sources in the catalog with secure spectroscopic redshifts.
\footnote{The multi\textendash wavelength XMM\textendash COSMOS catalog can be retrieved from: http://www.mpe.mpg.de/XMMCosmos/xmm53 \_release/,  version $1^{\rm st}$ November 2011. This is an updated release of the catalog already published by Brusa et al. (2010).}.

\subsection{Type1 and Type2 AGN samples}
\label{Type-1 and Type-2 AGN samples}

For the Type-1 AGN sample we have selected 1375 X\textendash ray sources detected in the [0.5-2]~keV band at a flux larger than $5\times10^{-16}$ erg s$-1$ cm$^{-2}$ (see \citealt{brusa10}). In this study we consider 426 objects from this sample that are spectroscopically classified as broad-line AGN on the basis of broad emission lines (${\rm  FWHM} > 2000 \,{\rm km \; s^{-1}}$) in their optical spectra (see \citealt{lilly07,trump09}).
The origin of spectroscopic redshifts for the 426 sources is as follows: 118 objects from the SDSS archive (\citealt{adelmanmccarthy05,kauffmann03}), 55 from MMT observations (\citealt{prescott06}), 66 from the IMACS observation campaign (\citealt{trump07}), 130 from the zCOSMOS bright $20$k sample, 49 from the zCOSMOS faint catalog (see \citealt{lilly07}), 
and 8 from individual Keck runs \citep{brusa10}. 
For a detailed description of the sample properties see \citet{2012ApJ...759....6E} and \citet{2012arXiv1210.3033H}.

To select an obscured AGN sample we have instead considered [2-10]~keV detected objects in the XMM--COSMOS catalog having fluxes larger than $3\times10^{-15}$ erg s$^{-1}$ cm$^{-2}$
(971 sources). From this sample we selected 288 AGN with spectroscopic redshift and optical spectra lacking broad emission lines (i.e. FWHM$<$2000 km s$^{-1}$): 259 are objects with either
spectrally unresolved, high-ionization emission lines, exhibiting line ratios indicating AGN activity, or not detected high-ionization lines, where the observed spectral range does not allow to construct line diagnostics \citep{brusa10}.
Not all AGN selected via this criterion show emission lines, e.g. 29 are classified absorption-line
galaxies, i.e., sources consistent with a typical galaxy spectrum showing only absorption lines. Further details on the obscured AGN sample and on their properties are given by \cite{lusso12}

Altogether, the sample used for testing the capabilities of AGNfitter consists on 714 X--ray selected active galaxies optically classified into 426 Type1 and 288 Type2 AGN. Due to Malmquist bias and owing to the different selection criteria,
the redshift distributions of the two population are considerably different, so that Type1 AGN cover a redshift range between $0.10<z<4.26$ ($\rm\langle z_{T1} \rangle=1.64 $) while Type2s present redshifts in the range $0.05<z<3.52$ ($\rm\langle z_{T2}\rangle=0.85 $).

\subsection{Multi-wavelength coverage}
\label{Multi-wavelength coverage}

The catalog includes multi-wavelength data from far-infrared to hard
X-rays: \textit{Herschel} data at 160~$\mu$m and 100~$\mu$m
(\citealt{2011A&A...532A..90L}), 70 $\mu$m and 24 $\mu$m MIPS GO3 data
(\citealt{2009ApJ...703..222L}), IRAC flux densities
(\citealt{2010ApJ...709..644I}), near-infrared J UKIRT
(\citealt{2008yCat.2284....0C}), H-band \citep{2010ApJ...708..202M},
CFHT/K-band data (\citealt{mccraken08}), optical multiband photometry
(SDSS, Subaru, \citealt{capak07}), and near- and far-ultraviolet bands
with GALEX (\citealt{2007ApJS..172..468Z}).  The observations in the
optical-UV and near-infrared bands are not simultaneous, as they span
a time interval of about 5 years: 2001 (SDSS), 2004 (Subaru and CFHT)
and 2006 (IRAC).  In order to reduce possible variability effects, we
have selected the bands closest in time to the IRAC observations
(i.e., we excluded SDSS data, that in any case are less deep than
other data available in similar bands).

Galactic reddening has been taken into account: we used the selective attenuation of the stellar continuum $k(\lambda)$ taken from Table 11 of \cite{capak07}. Galactic extinction is estimated from \citet{schlegel98} for each object.

\subsection{Treatment of upper limits}
\label{Treatment of upper limits}

The COSMOS photometric catalogs do not provide forced photometry at
the location of a source in the bands for which the object flux lies
below the formal detection limit. Instead, only upper limits are
reported. These upper limits can nevertheless be highly informative,
and place important constraints on model parameters, but we must
consider how to implement these upper limits into our Gaussian
likelihood. The correct approach would obviously
be to perform forced photometry in all bands where a
formal detection is not provided \citep{lang14}, which would give us
a measurement and an error, with a large error reflecting the fact
that the measurement (a marginal or non-detection) is very
noisy. However, adding forced photometry to the entire XMM--COSMOS AGN sample
is beyond the scope of the present work, hence we adopt the
following crude approximation which nevertheless allows us to
incorporate the information provided by the COSMOS upper limits.  
For non-detections, where the COSMOS catalog reports $5\sigma$ flux upper
limits ($F_{+}$), AGNfitter creates a fictitious data point at a flux
level of $0.5 F_{+}$, and a fictitious symmetric
error bar of $\pm (0.5 F_{+})$.
In this way, upper limits can
be trivially incorporated into our Gaussian likelihood in
equation~4.  Because our ``upper limit'' region spans from zero to $F_{+}$, 
this approach allows our MCMC sampling to accept
all models with fluxes lying between zero and the upper limit ones.
The inclusion of upper/lower limits in this fashion is necessary,
since the information they provide can be very helpful for 
constraining the likelihood of some models.

\section{Results and discussion on the test sample}\label{sec:realdata}

\begin{figure*}
\includegraphics[width=0.85\linewidth]{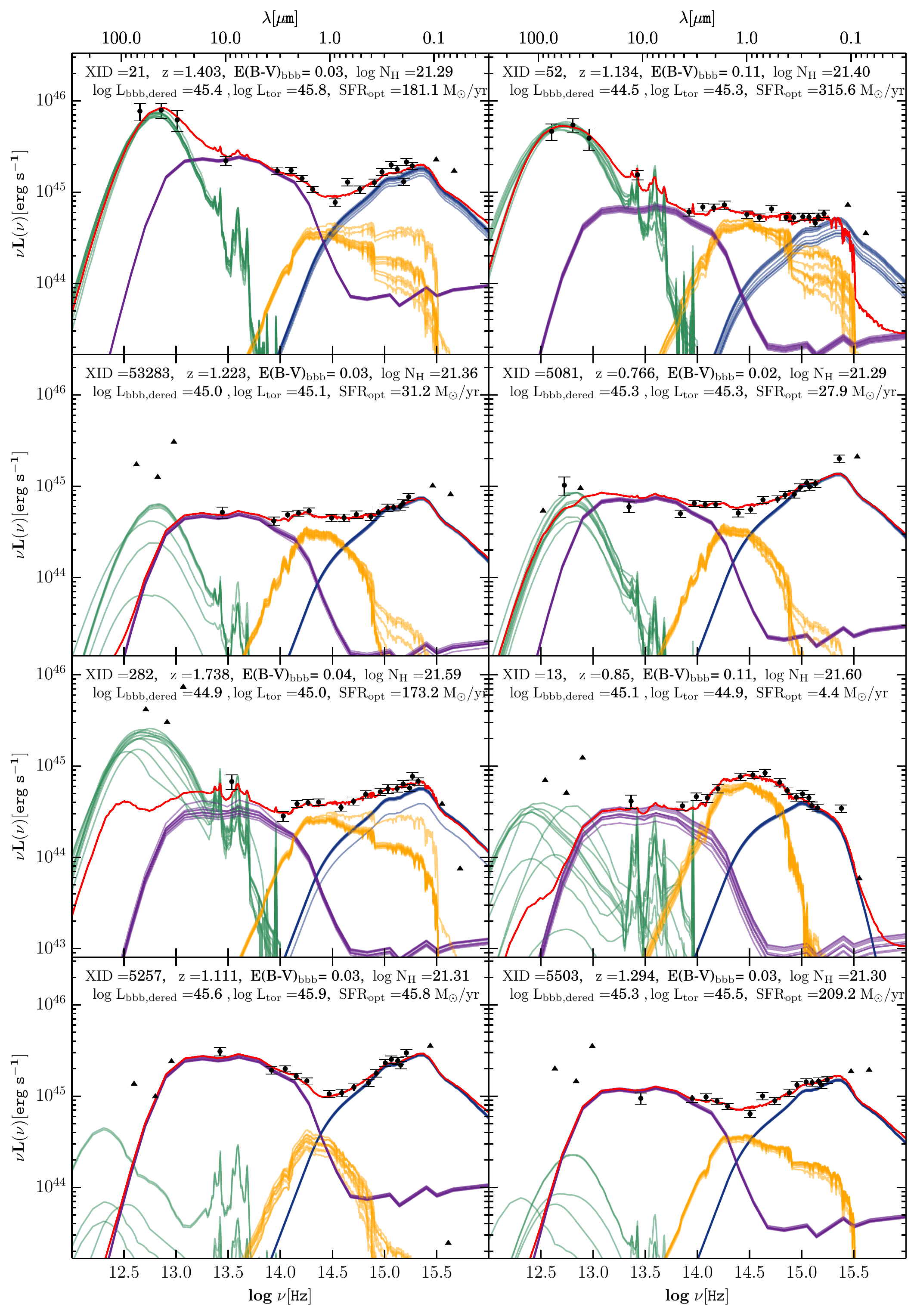}
\caption{Example SEDs for Type 1 AGNs: Photometric detections are plotted against rest-frame wavelengths/frequencies as circular markers with error bars, while non-detections are represented as triangles. SED shapes of the physical components are presented as solid lines with colours assigned similarly to Figure \ref{fig:templates}, while the linear combination of these, the 'total SED', is depicted as a red line. We pick 8 different realizations from the parameter posterior probability distributions  and over-plot these SEDs in order to visualize the effect of the  parameters' uncertainties on the SEDS.}
\label{fig:manySEDs-t1}
\end{figure*}

\begin{figure*}
\includegraphics[width=0.85\linewidth]{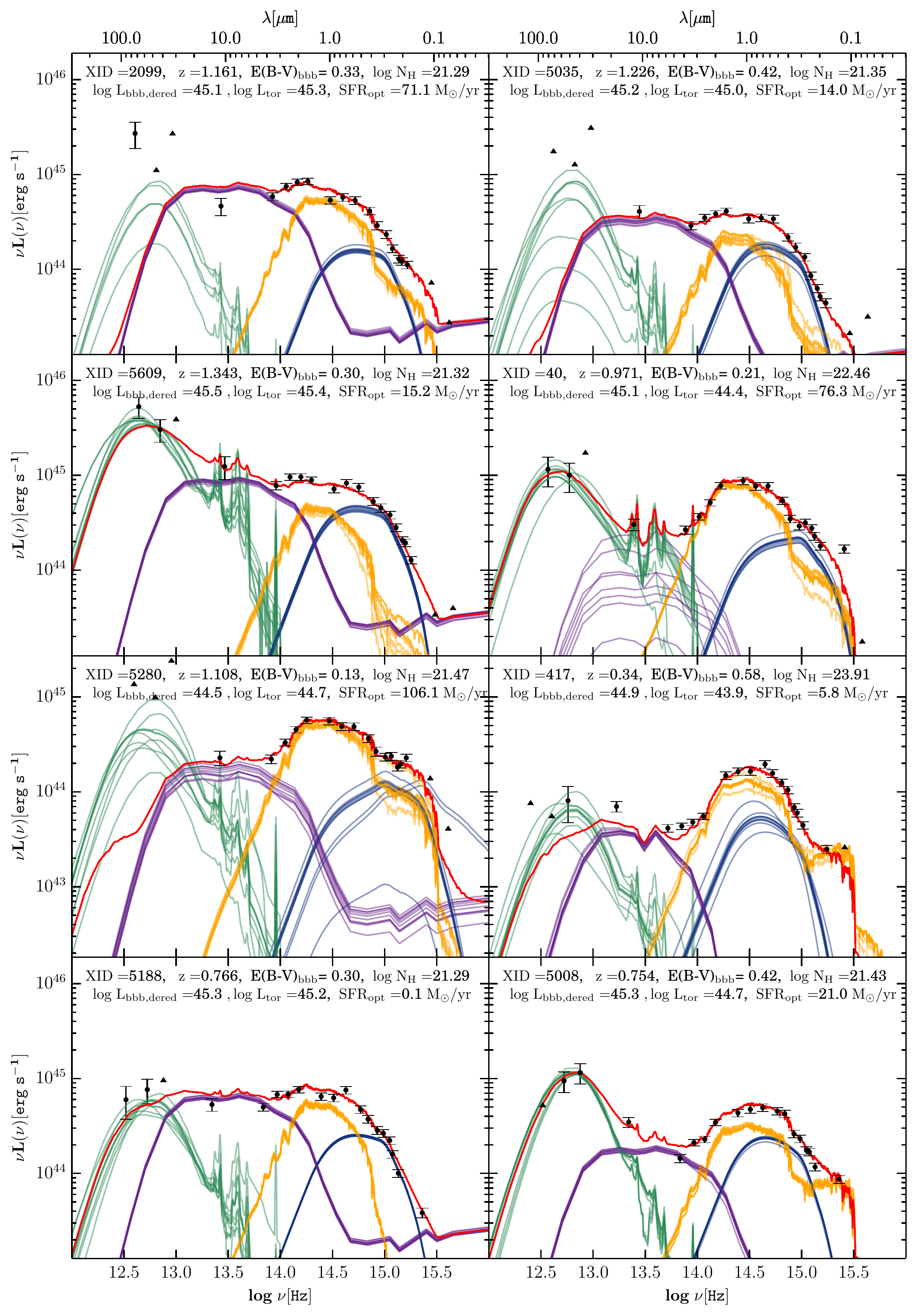}
\caption{Example SEDs for reddened Type 1 AGNs: Photometric detections are plotted against rest-frame wavelengths/frequencies as circular markers with error bars, while non-detections are represented as triangles. SED shapes of the physical components are presented as solid lines with colours assigned similarly to Figure \ref{fig:templates}, while the linear combination of these, the 'total SED', is depicted as a red line. We pick 8 different realizations from the parameter posterior probability distributions  and over-plot these SEDs in order to visualize the effect of the  parameters' uncertainties on the SEDS.}
\label{fig:manySEDs-t1red}
\end{figure*}

\begin{figure*}
\includegraphics[width=0.85\linewidth]{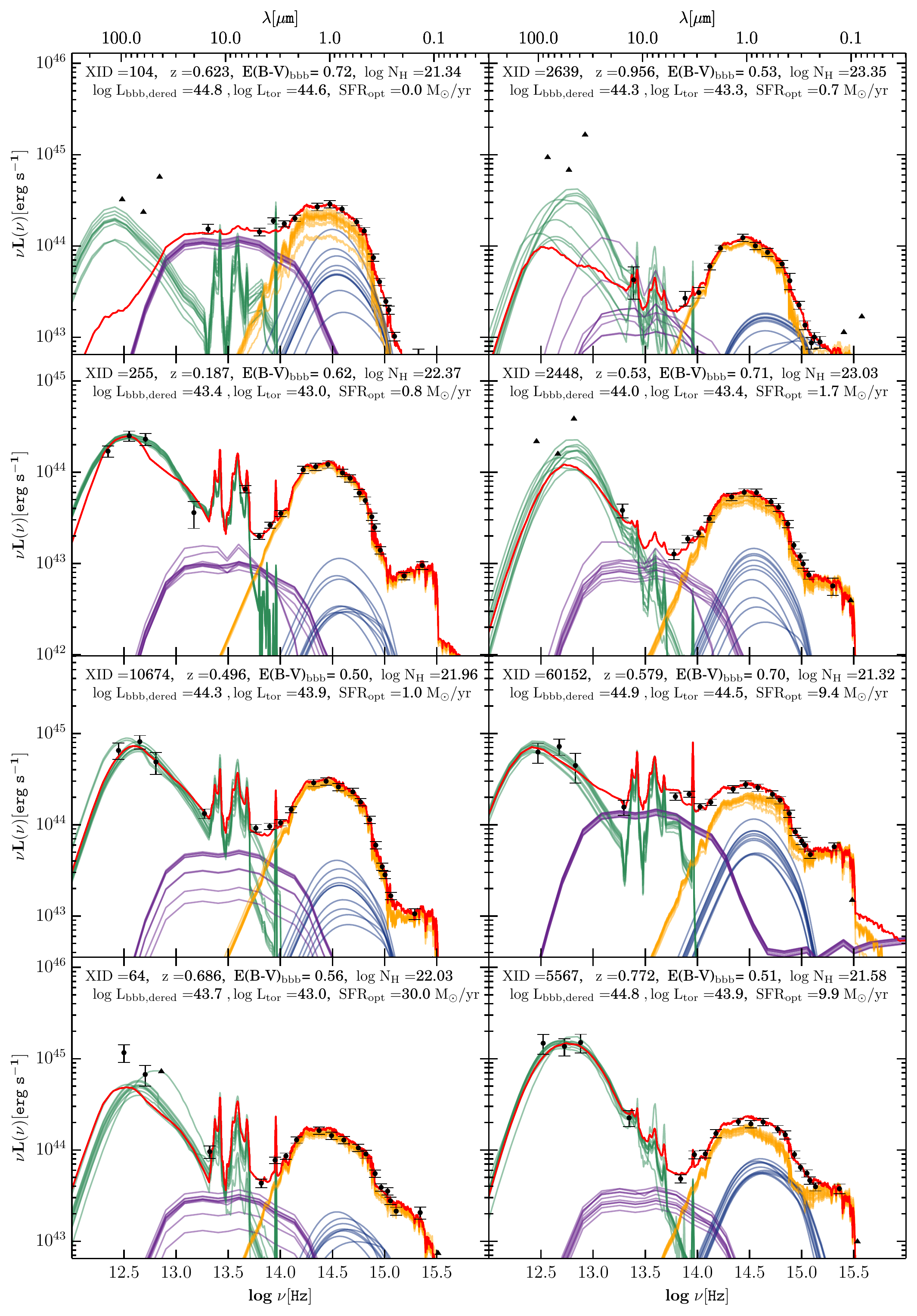}
\caption{Example SEDs for Type 2 AGNs: Photometric detections are plotted against rest-frame wavelengths/frequencies as circular markers with error bars, while non-detections are represented as triangles. SED shapes of the physical components are presented as solid lines with colours assigned similarly to Figure \ref{fig:templates}, while the linear combination of these, the 'total SED', is depicted as a red line. We pick 8 different realizations from the parameter posterior probability distributions  and over-plot these SEDs in order to visualize the effect of the  parameters' uncertainties on the SEDS.}
\label{fig:manySEDs-t2}
\end{figure*}

\subsection{Spectral energy distributions for Type1 and Type2 AGN}

In this section we discuss general properties within the Type1 and Type2 AGN populations from an SED perspective.
In Figure \ref{fig:manySEDs-t1} we present some examples of SED fitting for Type1 AGN having low levels of extinction in the optical-UV ($\rm E(B-V)_{\rm bbb}< 0.2$). To visualize the dynamic range of the parameter values included in the PDF, we randomly pick eight different realizations from the posterior PDFs of each corresponding component SEDs. 
Since the AGN sample we considered in our analysis is selected in the X-rays, it is less prone to be biased in the optical. This gives us the opportunity to test AGNfitter over a wider variety of AGN/galaxy contributions than optically selected ones.

XID=21 and 5257 are good examples of how an AGN SED looks like in the
case of both low-reddening and low-host galaxy contamination. Overall,
58\% ($248/426$) of the total Type1 AGN sample have $\rm E(B-V)_{\rm
  bbb}\lesssim 0.2$. The BBB and the torus properties for these
objects are nicely constrained and a clear dip at $1\mu$m is visible,
marking the division between hot-dust and accretion disc emission,
consistent with what it is typically found in optically selected AGN
samples
\citep{1994ApJS...95....1E,2006ApJS..166..470R,2011ApJS..196....2S}. On
the other hand, for these sources the fits to the the host galaxy
and cold-dust components
are not very informative. For the former this is because the AGN significantly outshines its host, whereas
the latter is simply due to a lack of FIR detections. 

There are also cases for which, although the BBB emission is
significant ($L_{\rm bbb,dered}>10^{45}$ erg s$^{-1}$), the galaxy
contribution at $\sim1\mu$m is not completely negligible, making the
overall shape of the AGN rather flat over several decades in frequency
(see XID=21, 52, and
5081 as examples).  As the BBB luminosity of the
source decreases ($L_{\rm bbb,dered}\simeq10^{44-45}$ erg s$^{-1}$),
even for cases with low BBB reddening, some sources require a 
significant galaxy contribution at $\sim1\mu$m (e.g. XID=13, 53283,
and 54513).  About 75\% of the sample have a contribution of the host
galaxy of about 50\% or more at $1\mu$m. The lack of a $1\mu$m
inflection point in the SED between the UV and near-IR bumps (i.e. a conspicuous
host galaxy contribution) is a common feature of X--ray selected AGN
samples, since these usually contain fainter AGN for which the contrast to galaxy emission is lower \citep{2012ApJ...759....6E,2014MNRAS.438.1288H}.

Figure~\ref{fig:manySEDs-t1red} presents some examples of Type1s with
significant amount of reddening in the optical-UV (E(B--V)$_{\rm
  bbb}\gtrsim0.2$; about 42\% of the Type1 AGN sample). The SEDs of
such sources (e.g. XID=2099, 5035, 5609, and 40) are characterised by
a sharp decline of the BBB  in the optical-UV ($\lambda \sim 0.5-0.2 \mu$m), and they can be
considered AGN of intermediate-type ($\sim$1.5 to 1.9).  Objects with
appreciable reddening still show degeneracy between BBB and the
host-galaxy, despite our implementation of a prior on the maximum
galaxy luminosity allowed at a given redshift. The BBB of a moderately
reddened Type 1 AGN can also be equivalently modelled with a bright
galaxy having very young stellar ages and a BBB with high reddening
values (e.g. see XID= 417, 5280 in
Figure~\ref{fig:manySEDs-t1red} as examples), and as expected this
degeneracy is even more pronounced in the case of low-luminosity AGN
($L_{\rm bol}\sim10^{44}$ erg s$^{-1}$).  To break this degeneracy, an
independent measurement of the host galaxy emission is needed,
although it is not trivial to estimate for unobscured AGNs. One
possibility is either to increase the threshold defining the
characteristic galaxy luminosity (as discussed in \S~\ref{Prior on the
  galaxy luminosity}), and/or to include spectral information. For
example, high \ion{H}{$\alpha$}/\ion{H}{$\beta$} ratio ($\gg3$)
indicates high reddening.

Regarding mid-IR properties
, Type1s are preferentially fitted with torus templates corresponding to an obscuring media of considerably low hydrogen column densities ($\log \NH < 22$). A quantitative study of the obscuration properties of this sub-sample is given in \S \ref{subsec:agnobscuration}.

Figure \ref{fig:manySEDs-t2} shows eight examples of SEDs for spectroscopically classified
Type2s.
Type2 AGN SEDs are dominated by the host galaxy emission at 0.5--2$\mu$m,
while the BBB component is obviously more difficult to constrain, spanning a wider dynamic range than in the Type1 cases.
Type2s are also characterised by high BBB reddening values (91\% have E(B--V)$_{\rm bbb}\gtrsim0.2$), as well as large column densities of the torus
component. Obscuration properties will be discussed in detail in \S~\ref{subsec:agnobscuration}.
Overall, the optical-UV portion of the Type-2 SEDs is very well fitted by stellar emission, with negligible contribution of the BBB. 
We also point out that the near-infrared emission at 10-20$\mu$m might be contaminated by emission features due to the stochastic heating of polycyclic aromatic hydrocarbon (PAH) molecules or carbon grains \citep[e.g.][]{flagey06}. 
These features are associated with massive star-forming regions at much lower dust temperature (T$<$100 K).
PAHs can be non negligible in obscured AGN and give rise to possible degeneracies with the torus emission produced by dust at pc scale. 
Such degeneracies may be solved by including priors on the cold-dust/torus emission which take into account the near-IR spectral information if available (e.g. line ratios, silicate absorption feature at 9.7 $\mu$m). 

It is reasonable to ask whether the physical properties inferred by AGNfitter allow us to robustly distinguish between unobscured and obscured AGN. In the following discussion, we will use the obscuration properties inferred by AGNfitter to re-classify the total sample into Type1s and Type2s and we will compare our classification to the spectroscopic one available in XMM--COSMOS.

\begin{figure}
\includegraphics[width=1\linewidth]{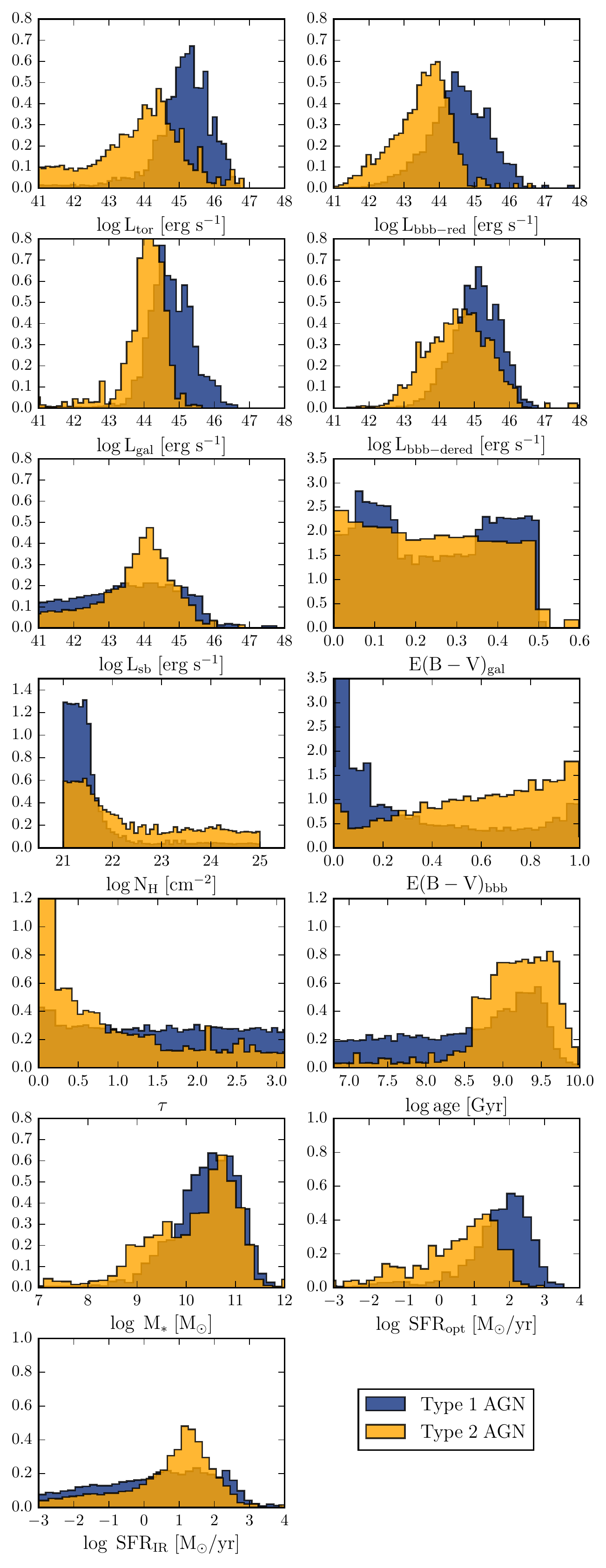}
\caption{Distribution of the output parameters for the whole Type1 (blue histrogram) and Type 2 (orange histogram) AGN population. These histograms were constructed sampling 500 random draws from each source's PDFs of each parameter.}
\label{fig:hist_par}
\end{figure}

\subsection{Physical parameters}

Figure \ref{fig:hist_par} present the output of AGNfitter on physical parameters for the total XMM--COSMOS AGN sample. As shown here, AGNfitter delivers vast information about integrated AGN luminosities, properties of the host galaxies and obscuration parameters. The large number of inferred properties can motivate a variety of science-cases related to AGN.

Figure~\ref{fig:hist_par} shows the parameter PDFs for the total Type1
(blue histograms) and Type2 AGN populations (orange histograms). These
histograms were constructed sampling 500 random draws from each
source's posterior PDF. Distinct from previous work, our treatment
thus enables a proper inclusion of the parameter uncertainties in the
histograms, allowing a fully probabilistic study of the global
properties of the two populations under the consideration, and other
inference effects which may shape the observed parameter PDFs.

The upper four histograms of Figure \ref{fig:hist_par} show the integrated luminosity distributions of Type1s and Type2s for $\rm \log L_{\rm tor}$, $\rm \log L_{\rm bbb-dered}$, $ \rm \log L_{\rm gal}$, and $\rm \log L_{\rm bbb-red}$. On average, Type1s appear to be more luminous than Type2s in all the four parameters, but this is clearly a selection effect as the average redshifts probed by these two samples are different, i.e. Type1s are on average observed at higher redshift than Type2s (see \S \ref{sec:dataset}). For instance, the $\rm \log L_{\rm tor}$ median values for Type1 AGN are on average over an order of magnitude higher ($45.20^{+0.61}_{-0.75}$) than those for Type2s ($44.01^{+0.98}_{-0.92}$). 

The distribution of $\rm L_{\rm bbb-red}$ for Type1 AGN has a median value
of $\rm \log L_{\rm bbb-red} = 44.55^{+0.83}_{-0.83}$ (the quoted uncertainties represent the 16th and the 84th percentiles of the distribution), while the median value for Type2s is $\rm \log L_{\rm bbb-red} = 43.59^{+0.62}_{-0.57}$, with a long tail to low BBB luminosities. Once we correct for extinction, the two $\rm\log L_{\rm bbb-dered}$ distributions nearly overlap within the uncertainties centered on similar median luminosity ranges: $45.08^{+0.65}_{-0.68}$ and $44.56^{+0.86}_{-0.61}$ for Type1s and Type2s, respectively, consistent with previous analysis on X--ray selected AGN samples \citep{lusso11,lusso12,bongiorno12,lusso13}. 
$58\%$ of the Type1 AGN appear to have relatively low levels of
extinction ( $\rm E(B-V)_{\rm bbb}<0.2$),
while the majority of Type2s exhibit significant reddening (
$91\%$ with $\rm E(B-V)_{\rm bbb}\geq0.2$), meaning a more appreciable reddening correction for the latter with respect to the Type1s.  As a result, the shift of the $\rm \log L_{\rm bbb-dered}$ distribution for Type2s to higher values is more pronounced than for Type1s. AGN obscuration properties related to the hydrogen column density parameter $N_{\rm H}$ will be further discussed in \S~\ref{subsec:agnobscuration}.

On the other hand, the starburst luminosities ($\rm \log L_{\rm sb,1-40~\mu m}$) show a rather flat and uninformative distribution for Type1 AGN with respect to Type2s. A similar result is obtained for the FIR star-formation rates ($\rm SFR_{\rm IR}$) (which are derived from the total infrared emission $\rm L_{\rm IR,~8-1000~\mu m}$). This is mainly due to the higher fraction of Herschel non-detections in the Type1 AGN with respect to the obscured
AGN population. Only $\sim$10\% of Type1 AGN are detected at both 100 and 160 $\mu$m, while Type2s have a detection fraction about 3 times higher (whereas at 70 $\mu$m$,\sim 8\%$ and 15\% of Type1 and Type2 are detected respectively).
We find that the median $\rm L_{\rm sb}$ luminosities and star-formation rates for Type2 AGN are $\rm \log~L_{\rm sb}= 43.80^{+0.85}_{-1.14}$ and $\rm \log~SFR_{\rm IR}= 0.95^{+0.82}_{-1.12}$, in agreement with previous analysis \citep{lusso11,bongiorno12}.
We conclude that at least one detection in FIR bands is essential for any analysis of the cold dust properties of the galaxy. Such a study will be presented for a sample of optically selected AGN and quiescent galaxies in a forthcoming paper (Calistro Rivera et al. in preparation).

Previous studies found X--ray selected AGN to reside preferentially in massive ($\rm M_\ast>10^{10}M_\odot$) bulge-dominated galaxies with red colours (e.g. \citealt{lusso11,bongiorno12}, but see also \cite{silverman09}; \cite{xue10}).
On average, stellar masses for Type1 AGN appear to be slightly larger than that of Type2s, although their difference is not statistically significant within their uncertainties ($\rm  \log M_\ast = 10.46 ^{+0.56}_{-0.73}$ and $\rm \log M_\ast =10.29 ^{+0.63}_{-0.82}$ for Type1s and Type2s, respectively). Due to their galaxy dominated SED shapes, Type2s have really well constrained host-galaxy physical parameters, showing relatively older stellar populations ($ age= 1.51 ^{+2.69}_{-0.80}$ Gyr) with respect to the Type1s ($ age = 0.27 ^{+2.02}_{-0.264}$ Gyr), in agreement with earlier works (\citealt{lusso11,bongiorno12}). Type1s have instead a much wider distribution of both $\tau$ and ages, which means that recovering the host-galaxy properties for unobscured AGN is a challenging task, subject to considerably larger uncertainties than for Type2s. In general, stellar masses are robustly constrained despite of the presence of an AGN, since these are primarily determined by the emission at around rest-frame 1$\mu$m, where the contrast between the BBB and galaxy is usually minimized. This is however not the case for the SFR parameter, since it is highly sensitive to the optical/UV SED, which is on its turn highly sensitive to AGN contribution, especially for Type1 AGN.

For completeness, we also reported in Figure~\ref{fig:hist_par} the distributions for the host-galaxy reddening $\rm E(B-V)_{\rm gal}$, and optical star formation rates ($\rm SFR_{\rm opt}$, corrected for extinction) for both AGN populations, although we note that the reliability of UV-based SFR indicators in galaxies has long been debated (e.g. \citealt{2000ApJ...544..218A,2002ApJ...577..150B}).

All in all, due to the different redshift ranges covered by the two AGN populations, ($\langle \rm z_{\rm T2}\rangle \sim0.85$, $\langle \rm z_{\rm T1}\rangle\sim1.64$), one should consider the effect of the redshift while interpreting the results to avoid comparing sources that may be at different evolutionary stages. Nonetheless, our main aim here is to provide an overall view of the full range of the physical parameters estimated by AGNfitter for the AGN population as a whole.

\subsection{AGN obscuration and classification}
\label{subsec:agnobscuration}
The parameters directly related to the obscuration properties of the nuclear emission are the dust reddening parameter $\rm E(B-V)_{\rm bbb}$ and the column density $\NH$. The parameter E(B-V)$_{\rm bbb}$
quantifies the absorption and reprocessing of the direct AGN emission by gas and dust along the line of sight at host-galaxy scales, while the column density parameter ($\log\NH$) describes the absorbing dust distributed mainly at nuclear scales.

We compare AGNfitter results on these parameters with those yielded by the non-Bayesian SED fitting code presented in \cite{lusso13} for the Type1 population. While they find a median dust reddening value of $\rm E(B-V)_{\rm bbb}\leq$0.03), 
AGNfitter finds a median value of $\rm E(B-V)_{\rm bbb}=0.13^{+0.58}_{-0.11}$ (see Figure \ref{fig:hist_par}).
The deviation from the literature results can be explained through the difference between our estimated median values from the PDFs  and the maximum likelihood ones (best-fit) usually adopted in the literature.

\begin{figure*}[t]
  \centering
  \includegraphics[width=\linewidth]{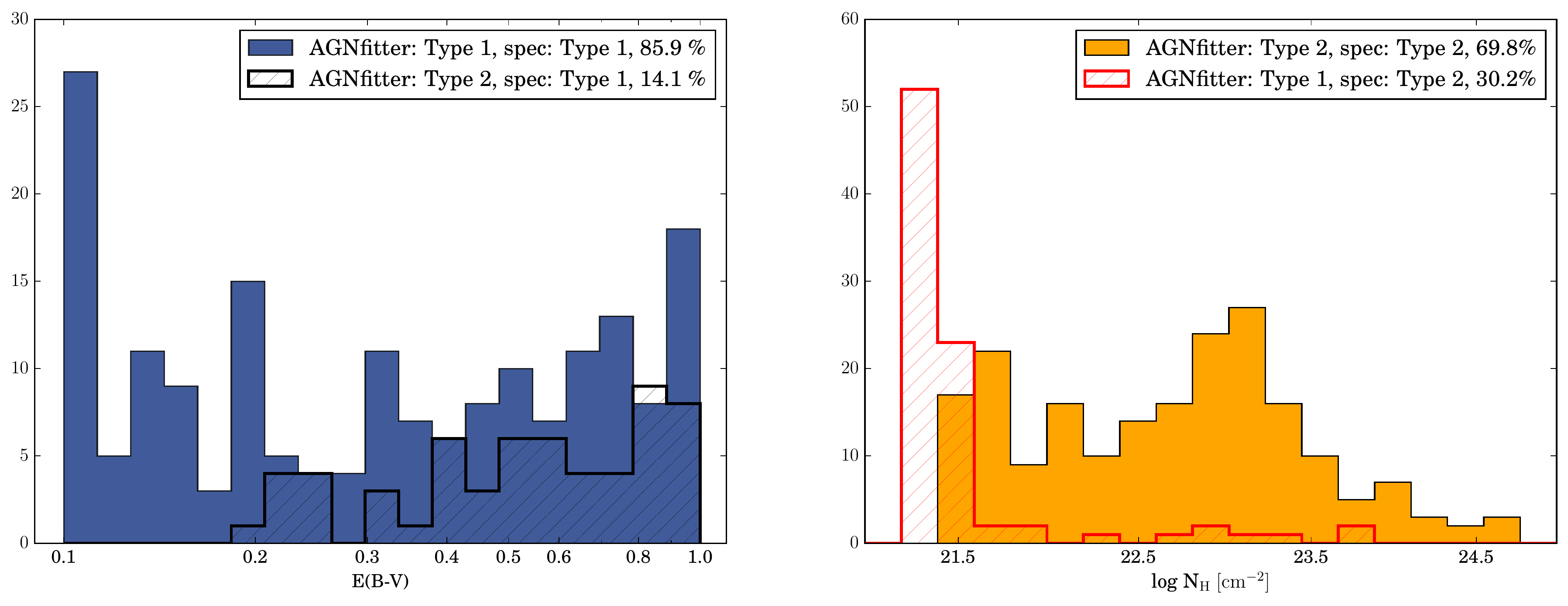}
  \caption{AGNfitter vs spectroscopic classification: the \textit{left panel} shows the distribution of the Type1 sources for the reddening parameter $\rm E(B-V)_{bbb}$. The filled blue distribution represents the spectroscopically classified Type1s that are also classified as such in AGNfitter, while the dashed black distribution represents the ones misclassified as Type2s. The \textit{right panel} shows the distribution of the Type2 sources for the column density parameters $\log \NH $. The filled orange distribution represents the spectroscopically classified Type2s that are also classified as such in AGNfitter, while the dashed red distribution represent spectroscopic Type2s misclassified as Type1s by AGNfitter. }
  \label{fig:classification}
\end{figure*}

This difference is especially relevant in the cases where the parameter $\rm E(B-V)_{\rm bbb}$ has an almost flat posterior PDF (as is the case for 36\% of the Type1 AGN, which have a total error in $\rm E(B-V)_{bbb}$ larger than 0.2).
For this kind of Type 1 AGN, the maximum likelihood estimate of the reddening parameter is not precise and the flat and broad
PDF push the median value to larger reddening values, which also results in large uncertainties on this parameter (see \citealt{assef10} for similar conclusions).

In contrast to the Type1s, no more than 9\% of the Type2 sub-sample present low reddening values of E(B--V)$_{\rm bbb}\leq$0.1. The large majority of Type2 are highly reddened sources with a population median value of E(B--V)$_{\rm bbb}=0.64^{+0.28}_{-0.25}$. 

A similar behaviour can be observed in the results for the column density parameter, $\log \NH$. As can be seen in Figure \ref{fig:hist_par}, Type1 AGNs clearly prefer low column densities with $\sim$ 86 \% having $\log \NH < 22$ as expected from their unobscured emission lines (population median value $\NH = 21.39 ^{+0.51}_{-0.26}$) . 
Type2 AGNs on the contrary, are more spread along large regimes of low and high column density, with only $\sim$ 45 \% having $\log \NH < 22$ (population median value $\NH = 22.02 ^{+1.95}_{-0.58}$).  
These column density values result from the different shapes of the torus SEDs, where higher column density is characterised by less near-IR emission of shorter wavelength, probably due to self-absorption of this energetic dust emission by the dust at lower temperatures distributed at larger scales (e.g. Figure \ref{fig:templates}).
Overall, AGNfitter is able to recover general obscuration properties of both AGN classes.

The slight bimodality in the distributions for these two independent
obscuration parameters provides the opportunity to develop a
classification strategy. A multiwavelength classification through
these AGNfitter parameters provides the clear advantage of being
sensitive to photometric measurements over a wide wavelength range
(mid-IR to opt-UV), in contrast to a spectral classification, which
covers only a few thousand Angstroms in the rest-frame. 
Moreover, while spectroscopy can be rather expensive and is not always available for all sources, our method can be easily applied to complete multi-wavelength datasets.

We define the classification of the two AGN populations to be the following:
\begin{itemize}

\item[] Type1 AGN:

\begin{itemize}
\item[] $\rm N_{\rm H}<21.5$ $\cup$ $0<\rm E(B-V)_{\rm bbb}< 0.2$
\end{itemize}
 
\item[] Type2 AGN:

\begin{itemize}
\item[] $\rm N_{\rm H}>21.5$ $\cap$ $\rm E(B-V)_{\rm bbb}> 0.2$
\end{itemize}

\end{itemize}

Although this is a hybrid sample because of the different selection effects employed,
it nevertheless provides useful estimates for the true completeness and efficiency of classification for  more uniformly selected X-ray AGN samples.
The results of our classification scheme are presented in Fig. \ref{fig:classification}.
As can be seen in the left panel,  the majority of the broad emission line
AGNs spectroscopically classified as Type1s are also classified as
Type1s by the AGNfitter obscuration parameters, proving a completeness of (366/426, $\sim$ 86\%).
For the case of Type2s in the right panel of Fig. \ref{fig:classification}, our classification method is complete in $\sim$ 70\% (201/288).
A small fraction of the spectroscopically classified Type1s are misclassified as Type2s by AGNfitter's method (89/455), showing to be efficient at $\sim$80\% for Type1 AGN. 
The false-positive ratio is slightly larger for Type2s, being 60/261 of spectroscopically classified Type2s missclassified as Type1 AGN by our method, proving an efficiency of  $\sim$77\% for Type2 sources.

The cases of disagreement with the spectroscopic approach does not necessarily imply that the SED fitting classification is incorrect. On the contrary, a multiwavelength approach as AGNfitter can be more sensitive to the global properties of the sources for such a classification, since it analyses the obscuration level in different independent components of the SED. 

A similar approach has been used by \cite{assef13} based on a non-Bayesian SED fitting code, where they adopt E(B −V) = 0.15 as the dividing line for their classification. 
They choose this classification using an estimate from the standard X-ray boundary of a gas
column density of $\rm N_{H} = 10^{22} \rm cm^{−2}$ \citep{ueda03} and the median value of the ratio $\rm E(B−V)/N_{\rm H} = 1.5\times 10^{−23} cm^2 mag$ observed for the sample presented in \cite{maiolino01}. 
Although a direct completeness comparison to the method of \cite{assef13} is not possible, since they do not test their method on spectroscopic classification, we calculate the completeness of their classification strategy applied to the fitting results of our sample.
Following \cite{assef13}  and using E(B −V) = 0.15 as the dividing line for the classification, we obtain a completeness ratio of 54\% and 91\% and an efficiency of 90\% and 57\% for Type1 AGN and Type2 AGN respectively. In general, our method present slightly better completeness and efficiency ratios, recovering a significantly larger fraction of Type1 sources, while slightly compromising on the completeness of Type2s. As this disagreement is likely to arise from the different SED-fitting approaches used to infer the obscuration parameters, a direct comparison of their classification strategy to spectroscopic classified samples would be needed.

\section{Summary and Conclusions}

We introduced AGNfitter: a fully Bayesian statistical tool for fitting
and decomposing SEDs of active galaxies through MCMC sampling of the
model parameters. The total active galaxy model in AGNfitter
consists of the host galaxy emission, modelled as a combination of a
stellar component and a starburst cold gas component and the nuclear
AGN emission, modelled as a combination of an accretion disk (BBB)
and a hot dust torus component.  For both AGN and host galaxy
models in the optical and UV, the effect of reddening along the line
of sight is accounted for and corrected. Through informed sampling,
the algorithm explores the parameter space that defines this model and
computes the full PDFs
of the
parameters. AGNfitter calculates the median 
values with
respective uncertainties of the following parameters:
\begin{itemize}
\item physical parameters: $\tau$,  $age$, $\log N_{\rm H}$,
\item reddening parameters: E(B--V)$_{\rm bbb}$ and E(B--V)$_{\rm gal}$,
\item normalization variables: SB, BB, GA, TO
\end{itemize}
Multiple relevant properties for the characterization of the AGN and the host galaxy are also calculated. These are, for example, the integrated luminosities of the AGN components (e.g. L$_{\rm bol}$, L$_{\rm bbb}$, L$_{\rm bbb-dered}$ and L$_{\rm tor}$) and host galaxy properties (e.g. L$_{\rm gal}$, L$_{\rm sb}$, $M_\ast$, SFR$_{\rm opt}$, and SFR$_{\rm FIR}$).

The capability of AGNfitter in properly inferring these source properties was tested on mock active galaxies created as prototypical Type1 and Type2 SEDs. 
AGNfitter could accurately recover the input parameters of the synthetic SEDs and also provide insights into the degeneracies between emission components through the shapes of their PDFs.

The performance of the code for real data was tested
on a sample of X-ray selected AGN from the
XMM--COSMOS survey, which provides 15 band photometry from 
the UV to the far-IR for the construction of the SEDs.
AGNfitter was
applied to 714 sources (426 Type1 and 288 Type2 AGN),
which were previously spectroscopically classified as Type1 and Type2
sources by their optical emission lines.  The fitting-results include two independent model parameters,
which are proxies for AGN obscuration.
This
allowed us to develop a classification strategy for unobscured
(Type1) versus obscured (Type2) AGN,
which shows a
great agreement with the spectroscopic classification. The completeness fraction of our classification scheme is $\sim
86\%$ and $\sim 70\%$, with an efficiency of $\sim
80\%$ and $\sim 77\%$, for Type1 and Type2 AGNs respectively.

The complexity of AGN physics leaves many possibilities of improving
the physical models used to approximate AGN emission. While they might
be more detailed and accurate models for the different AGN components,
these imply the treatment of larger parameter spaces with an unavoidable increase of degeneracies among the parameters. 
In those cases, the use of a Bayesian methodology as the one of AGNfitter is even more crucial.

The efficiency and code structure of AGNfitter allows new models to be easily integrated to the existing libraries. One significant improvement would be to implement a more flexible BBB model, which can be based on a physically motivated accretion disc model, possibly dependent on black hole mass and disc accretion rate \citep[e.g.][]{slone12}, instead of a single empirical template. A further possible change could be the implementation of a more complex prior on the galaxy models to reflect the information from their total luminosity function. An interesting addition would be to allow the user to explore the effect on the galaxy/AGN parameters if an energy balance between the cold-dust and optical-UV star-formation is considered \citep[e.g.][]{2008MNRAS.388.1595D}. Moreover, the implementation of a model of the IGM absorption such that rest-frame UV wavelengths at $\lambda<1250$\AA\ can be included in the fitting procedure, would be an additional improvement of our model.

AGNfitter's multi-wavelength approach allows the simultaneous study of multiple physical processes. This versatility makes of AGNfitter a flexible tool to address several physical questions related to AGN and galaxies. The AGNfitter python code is publicily available at \url{https://github.com/GabrielaCR/AGNfitter}.

\section*{Acknowledgements}

The authors thank the anonymous referee for a careful reading and useful comments which have improved this manuscript.
Joseph F. Hennawi acknowledges generous support from
the Alexander von Humboldt foundation in the context
of the Sofja Kovalevskaja Award. The Humboldt foundation
is funded by the German Federal Ministry for Education
and Research. 
Gabriela Calistro Rivera gratefully acknowledges support from the European Research Council
 under the European Union’s Seventh Framework Programme (FP/2007- 2013)/ERC Advanced Grant NEWCLUSTERS-321271.
We thank the members of the ENIGMA group \footnote{http://www.mpia-hd.mpg.de/ENIGMA/} at the
Max Planck Institute for Astronomy (MPIA) for insightful suggestions and discussions. 
We thank Daniel Foreman Mackey for his support 
on the implementation of the code Emcee as the MCMC core of AGNfitter.
We thank Neil H.M. Crighton, Jose Onorbe, Alberto Rorai, Kenneth Duncan, and Marco Velliscig for important contributions in the scripting of the algorithm in Python.  
We thank Micol Bolzonella and Elena Zucca for useful comments and suggestions on the implementation of the prior on the galaxy luminosity function.
We thank Huub R\"ottgering for his support during the final phase of the project.

\bibliographystyle{apj}
\bibliography{PAPER2_arxiv}

\end{document}